\documentclass[journal ]{new-aiaa}
\usepackage[utf8]{inputenc}
\usepackage{textcomp}
\usepackage{hyperref}
\usepackage{siunitx}
\usepackage{xcolor}
\usepackage{subcaption}
\usepackage{lineno}
\usepackage{tikz}
\usepackage{graphicx}
\graphicspath{{./figures/}}
\usepackage{cleveref}
\usepackage{amsmath}
\usepackage[version=4]{mhchem}
\usepackage{siunitx}
\usepackage{longtable,tabularx}
\usepackage{multicol}
\usepackage{multirow}
\usepackage{array}
\newcolumntype{P}[1]{>{\centering\arraybackslash}p{#1}}
\newcolumntype{M}[1]{>{\centering\arraybackslash}m{#1}}

\title{Aerodynamic Characterization of a Fan Array Wind Generator}

\author{Songqi Li \footnote{Postdoctoral Research Associate, School of Mechanical Engineering and Automation, AIAA Member, \href{mailto:lisongqi@hit.edu.cn}{lisongqi@hit.edu.cn}.},  
	Yutong Liu\footnote{Graduate Student, School of Mechanical Engineering and Automation, \href{mailto:liuyt0318@163.com}{liuyt0318@163.com}.},  
	Zhutao Jiang\footnote{Graduate Student, School of Mechanical Engineering and Automation, \href{mailto:jiangzhutao123@stu.hit.edu.cn}{jiangzhutao123@stu.hit.edu.cn}.},
	Gang Hu\footnote{Professor, School of Civil and Environmental Engineering, \href{mailto:hugang@hit.edu.cn}{hugang@hit.edu.cn}.},
	and 
	Bernd R. Noack \footnote{National Talent Professor, School of Mechanical Engineering and Automation, \href{mailto:bernd.noack@hit.edu.cn}{bernd.noack@hit.edu.cn}.}}
\affil{Harbin Institute of Technology, 518055 Shenzhen, People's Republic of China}
\author{Franz Raps\footnote{Chair Professor, College of Urban Transportation and Logistics, \href{mailto:franzraps@sztu.edu.cn}{franzraps@sztu.edu.cn}.}}
\affil{Shenzhen Technology University, 518118 Shenzhen, People's Republic of China}

\begin{document}

\maketitle
\begin{abstract}
Experimental assessment of safe and precise flight control algorithms 
for unmanned aerial vehicles (UAVs) under gusty wind conditions
requires the capability to generate a large range of velocity profiles.
In this study, 
we employ a small fan array wind generator
which can generate flows with large spatial and temporal variability.
We perform a thorough aerodynamic characterization operating the fans uniformly from a low to the maximum level.
PIV and hot-wire measurements indicate a jet-like flow 
with nearly uniform core which monotonously contracts in streamwise direction
and surrounding growing unsteady shear-layers.
These complex dynamics results in a limited region with desired flow profile and turbulence level.
The experimental results shed light on the flow generated by a full-scale fan array wind generator, 
and indicate the need for further improvements via properly designed add-ons and dedicated control algorithms. 
\end{abstract}

\section{Introduction}
\lettrine{T}{he} rapid development and increased application of unmanned aerial vehicles (UAVs) 
and electrical Vertical Take-Off and Landing aircrafts (eVTOL)
demand high levels of agility and precise control \citep{Ducard_2021,Kim_2019,Liu_2017,Amin_2016,Jasim_2020}. 
For instance, air taxis must safely land on rooftops and vertiports \citep{Rajendran_2020}, 
cargo delivery drones must accurately follow the planned trajectories \citep{RojasViloria_2020}, 
and rescue drones require precise control in extreme weather conditions \citep{Ol_2008,MohdDaud_2022}. 
However, the atmospheric boundary layer (ABL) may pose a considerable challenge
to the safe and precise control of UAVs due to its turbulent, nonlinear, and unpredictable nature \citep{Garratt_1994,Barlow_2014}.
The development and application of more advanced flight control algorithms become critical for UAVs to fight against environmental disturbances. 
To test the performance of flight control algorithms, 
it is necessary to simulate artificial winds with temporal and spatial variability during the flight tests \citep{Noca_2019,Azzam_2023}.

Multiple experimental techniques have been developed to generate artificial winds with spatial and temporal variability.
Roughness elements are the most commonly used device to simulate ABL profiles in wind tunnels \citep{Farell_1999,Tieleman_2003,Zhao_2022}.
To generate flow unsteadiness, 
deploying louvers with multiple variable blades or vanes \citep{Charnay_1976,Greenblatt_2016,Haan_2006}
is the most popular strategy.
Other artificial disturbances in the flow, such as 
wavy walls \citep{Holmes_1973},
rotating cylinders \citep{Tang_1996},
and a pitching/plunging airfoil \citep{Wei_2019}
are also capable to generate unsteady incoming flow.
An alternative strategy employs  arrays of randomly
actuated jets  to generate desired turbulent profiles \cite{Variano2008,Bellani2013,Carter2016,PrezAlvarado2016,Esteban2019,Lawson2021}.
Pioneered by \citealt{Makita_1991}, active grids represent another type of technique to generate flow with spatial and temporal variations.
Active grids consist a series of small wings that are mounted on the horizontal and vertical shafts.
The rotation of shafts will alter the local blockage and gives rise to different incoming flow properties.
Although the idea of active grids is first proposed to generate turbulence with desired characteristics \citep{Mydlarski_1996,Mydlarski_1998,Kang_2003,Larssen2010,Hearst2015},
the capability of such devices to generate different flow profiles has recently been exploited and reported in \citep{Hearst_2017,Neuhaus_2021,Knebel_2011,Azzam_2023}
among many others.

The recent emergence of Fan Array Wind Tunnels (FAWTs) and Fan Array Wind Generators (FAWGs) represents a new opportunity to design artificial wind with specified spatial patterns and temporal profiles.
By placing an array of fans (or propellers) at the test section inlet (FAWT,\citep{nishi1997turbulence,Ozono_2006,Smith_2012,Wang_2018,Catarelli_2020}), or in an open environment (FAWG, \citep{Johnson_2009,Aly_2011,Noca_2019}), fan arrays are capable to generate customizable wind conditions by controlling the rotation speeds of fan elements.
As the fans can be individually controlled, FAWTs and FAWGs can generate wind profiles with rich spatial and temporal variability.
Since the operation of FAWGs doesn't rely on currently existing wind tunnels, wind generators quickly become a popular choice for UAV flight tests.
Recently, several drone flight experiments are conducted where FAWGs are employed to generate environmental disturbances \citep{Olejnik_2022,Guibert_2022,OConnell_2022,Walpen_2023}.
These encouraging results suggest the potential of FAWGs to  help test and improve the design of UAV control systems.
However, from the aerodynamical point of view, there is still a lack of a complete characterization of flow generated from FAWGs.
Although flow measurements from pointwise sensors have been reported in \citep{Noca_2021}, 
flow field measurements with satisfying spatial resolution are still necessary in order to fully understand the flow properties.
However, given the typical scale ($\mathcal{O}(\SI{1}{\meter})$) of FAWGs, it is difficult to directly apply Particle Image Velocimetry (PIV) for  wind generator flow measurements, given the restrictions in laser power, camera positioning, etc.

In this work, we design and construct a Small Fan Array Wind Generator (SFAWG), 
which has an overall size of $\SI{40}{\centi\meter}\times \SI{40}{\centi\meter}$ and contains 100 fan elements.
We select small fans to construct the wind generator, such that typical flow measurement tools in the laboratory become available to characterize the flow field from the small wind generator.
With flow measurements from SFAWG, we hope to infer the major flow characteristics of full scale wind generators.
We conduct a complete aerodynamical study of this facility, including hotwire scanning on four near field planes close to the fan exit, planar PIV measurements on a streamwise plane, and stereoscopic PIV measurements on two far field planes.
The discussion focuses on the characteristics of the streamwise velocity profiles when all fans are operated under the same rotation speeds, as the profiles of the streamwise velocity component are of the most importance regarding the generation of artificial wind gusts.
From the analysis of the experimental data, we observe complex flow dynamics, including the initial mixing in the near field, the stretched and distorted jet profiles in the far field, 
and the fast expansion of the annular shear layers.
These observations are crucial to (1) understand the complex flow dynamics in full-scale FAWGs, 
(2) guide the placement of UAVs in flight tests, and (3) develop add-on devices and control algorithms to further improve the flow quality.

The manuscript is structured as follows. 
In \Cref{sec:methodology}, we provide a detailed description of the experimental facility and techniques used in this study. 
This includes the design of the SFAWG, as well as the instrumentation used for flow measurements. 
\Cref{sec:results} presents the flow measurement results, including the mean flow statistics and an extended discussion on the turbulence characteristics.
we offer our major observations and provide an outlook for future research in \Cref{sec:conclusions}.

\section{Experiment Facility and Techniques}\label{sec:methodology}
This section outlines the design of  the Small Fan Array Wind Generator (SFAWG), 
and the experimental instrumentation for flow field characterization.
Featuring individual control of $10\times 10$ fan elements, SFAWG represents a miniatured version of full scaled wind generators. 
We can utilize this specialized test facility to study the flow generated from fan arrays, and to implement flow control algorithms that enable us to achieve desired flow properties.
This study focuses on the experimental characterization of SFAWG using hotwire anemometry (HWA) and particle image velocimetry (PIV). 
In \Cref{sec:design}, we present a detailed description of the experimental facility, including its design and construction. 
This is followed by a description of the experimental setup which is provided in \Cref{sec:config}.

\begin{figure}[!htb]
	\centering
	\begin{subfigure}{0.49\textwidth}
	\centering
		\includegraphics[width=.8\textwidth]{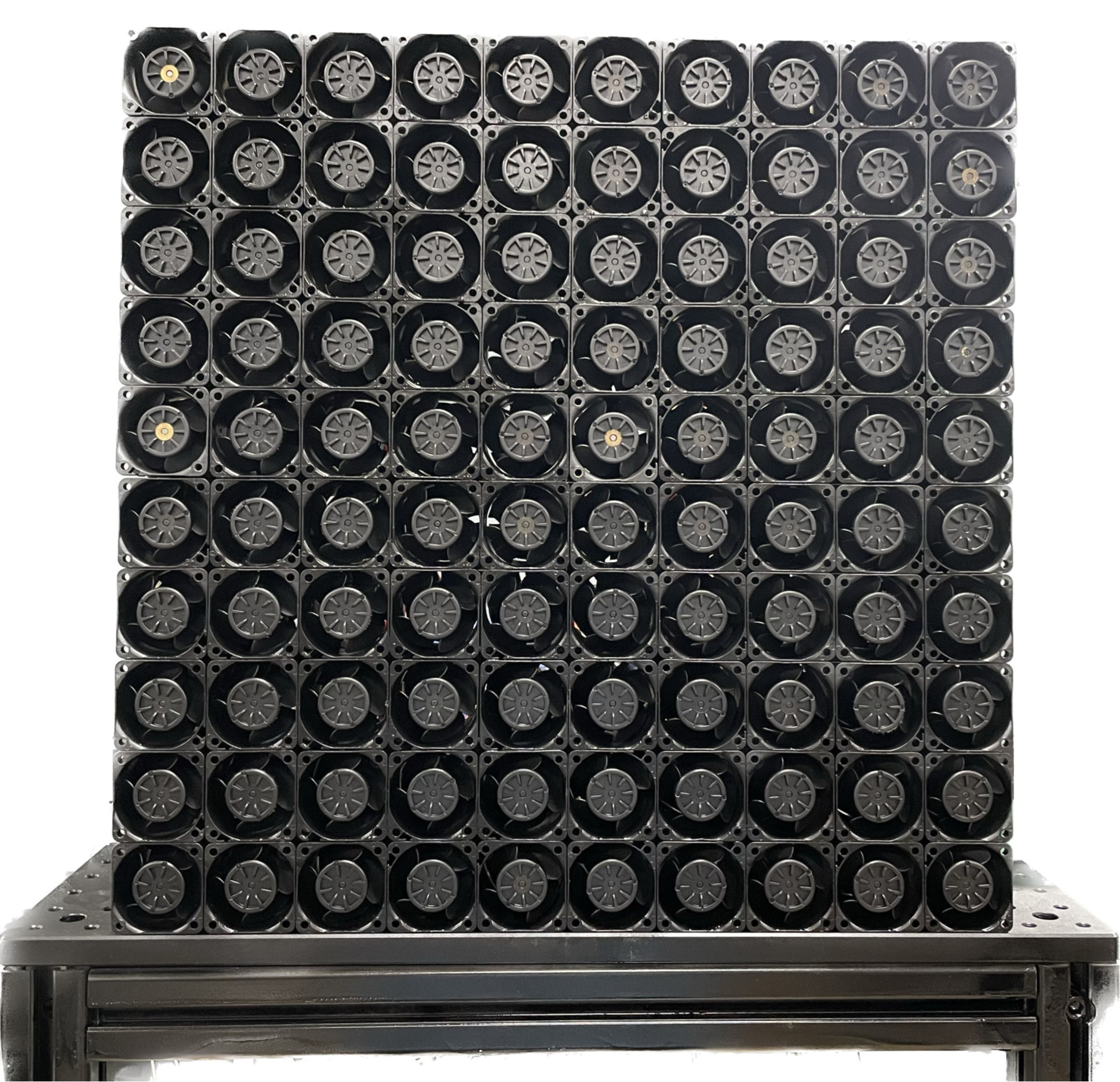}
		\caption{}
		\label{fig:W1_1}
	\end{subfigure}
	\begin{subfigure}{0.49\textwidth}
	\centering
		\includegraphics[width=.8\textwidth]{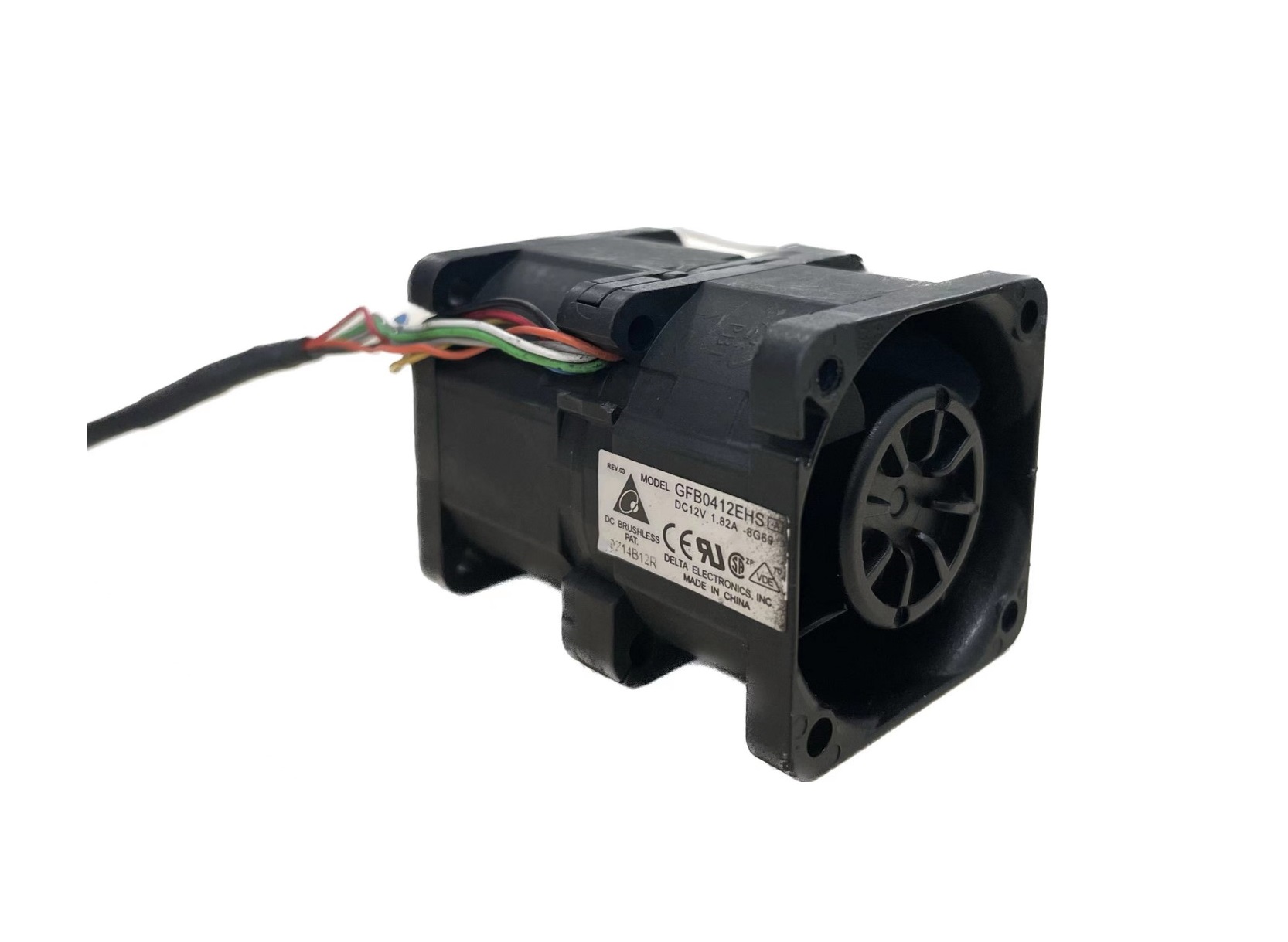}
		\caption{}
		\label{fig:W1_2}
	\end{subfigure}
	\\
	\begin{subfigure}{0.7\textwidth}
	\centering
		\includegraphics[width=.99\textwidth]{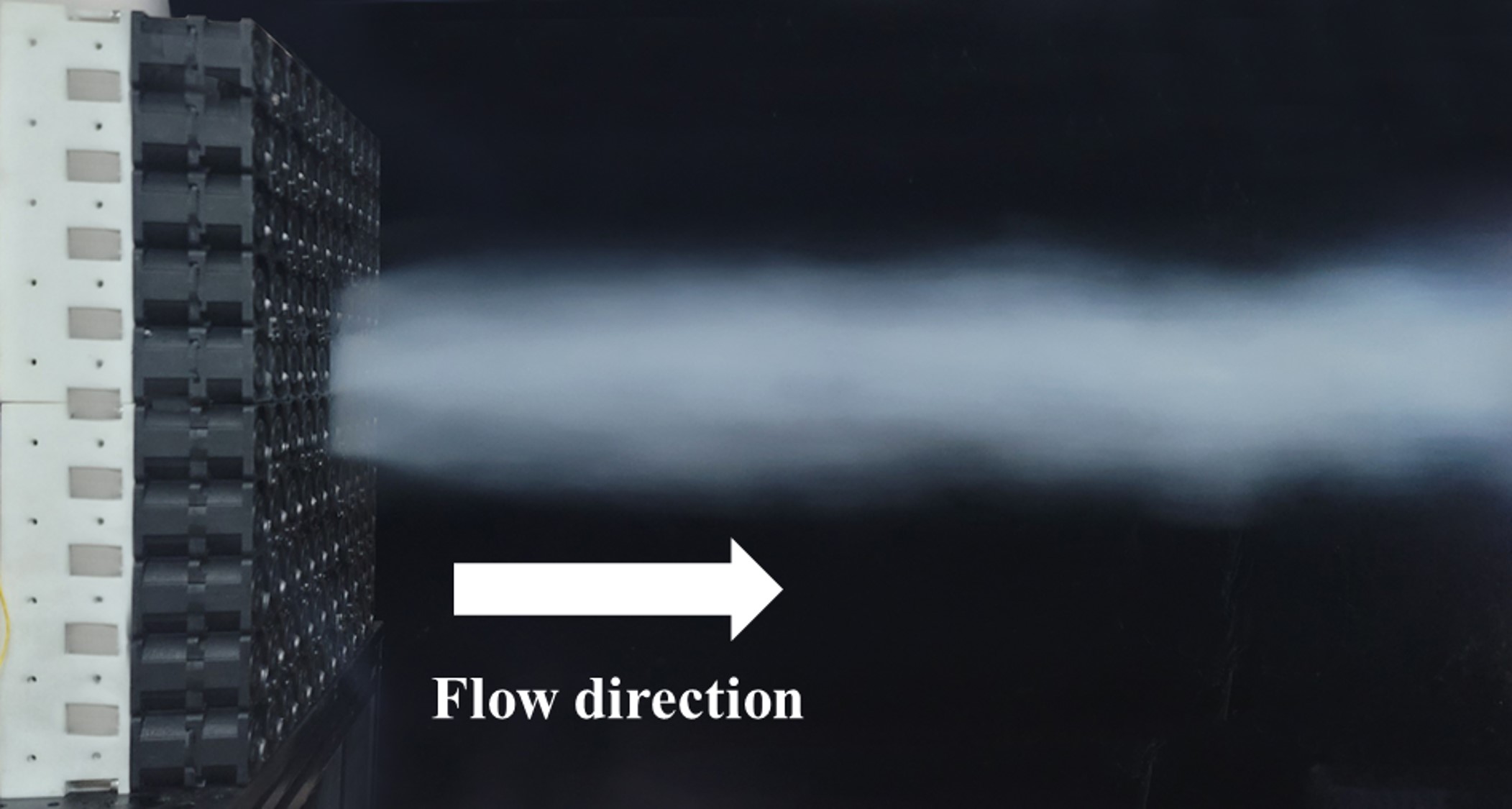}
		\caption{}
		\label{fig:W1_3}
	\end{subfigure}
	
	\caption{A graphical introduction of small fan-array wind generator (SFAWG). (a) front view of SFAWG; (b) the fan element used in the wind generator; (c) smoke visualization of wind generated from SFAWG.}
	\label{fig:W1}
\end{figure}

\subsection{Experiment Facility: Small Fan Array Wind Generator} \label{sec:design}
As show in \Cref{fig:W1_1}, the small fan array wind generator is composed of $10 \times 10$ axial fans.
The fans utilized in this study is Delta Electronics GFB0412EHS.
This dual-stage counter-rotating fan has a squared cross-section and a dimension of $\SI{40}{\milli\meter}\times\SI{40}{\milli\meter}\times\SI{56}{\milli\meter}$.
The blades on the front side rotate counterclockwise, while the blades on the rear side rotate clockwise.
We stack 100 fan elements in a nearly seamless manner both horizontally and vertically forming a $10\times 10$ fan array.
Specially designed frames and multi-hole washers are used to fix the fans, and the assembly is mounted on a stand that is $\SI{1.5}{\meter}$ tall and made of aluminum extrusions.
In this study, the wind generator is placed inside an empty chamber which is $\SI{8}{\meter}\times\SI{8}{\meter}\times\SI{3.5}{\meter}$.
We place the wind generator on one side of the chamber, and the air is blown towards the opposite end (\Cref{fig:W1_3}).
A cooling system is used inside the chamber to keep a constant ambient temperature of $\SI{23}{\degreeCelsius}$.

\begin{figure}[!htb]
	\centering
	\includegraphics[width = .9\textwidth]{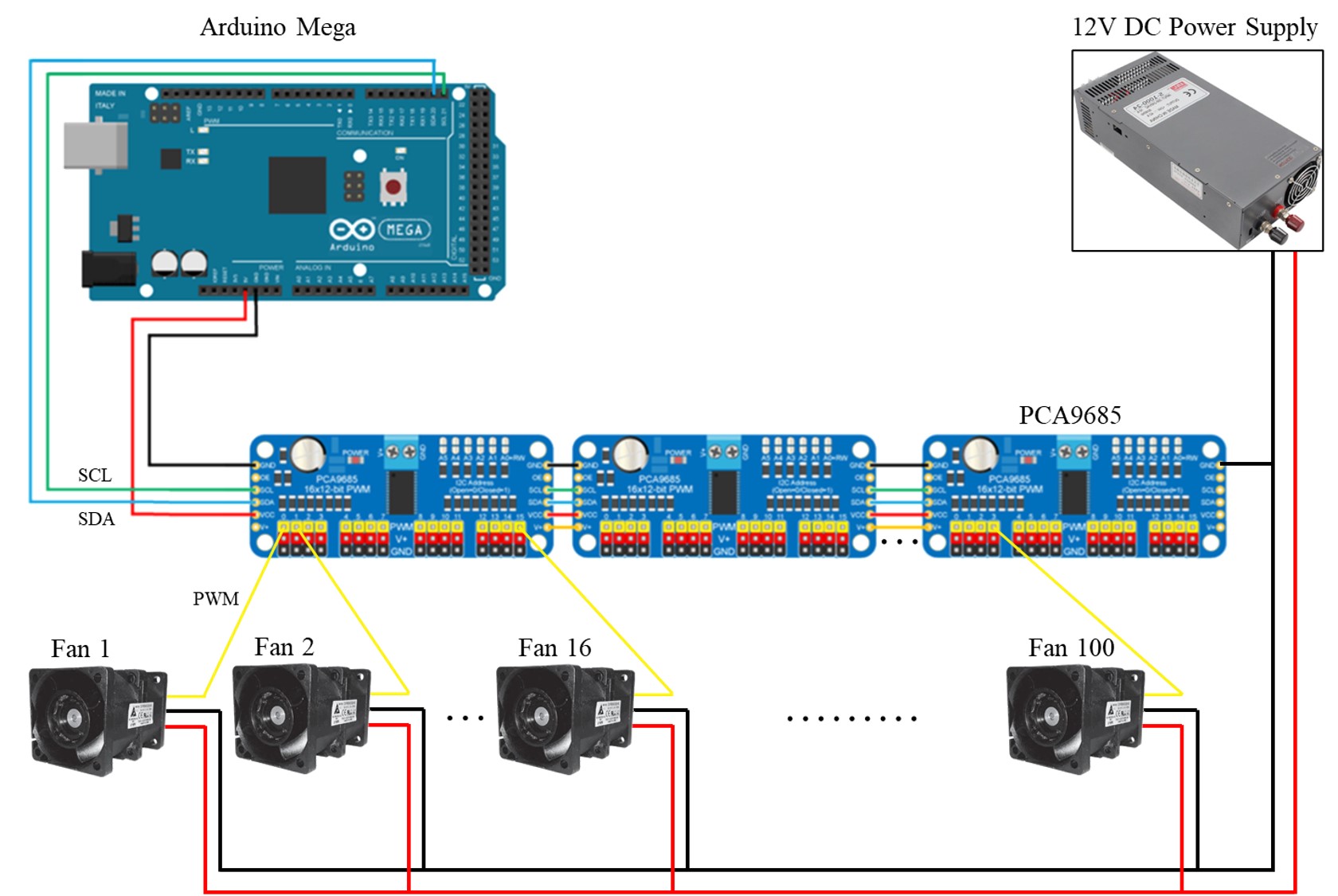}
	\caption{The wiring diagram for the individual control of fan elements in SFAWG.}
	\label{fig:Wiring}
\end{figure}

We connect the fan array to a 12 VDC power supply, and control 
the rotation speed of  each fan element using pulse-width-modulation (PWM) signals with variable duty cycles.
As the duty cycle increases, the fan speed will increase accordingly.
\Cref{fig:Wiring} presents the wiring diagram of the wind generator.
To provide 100-channel PWM signals to the fan array, we chose a series of 7 PCA9685 PWM drivers.
Each PWM driver is capable to generate 16-channel independent PWM outputs, and the duty cycle of every output is controllable.
An Arduino Mega is used as the host controller to communicate with the PWM drivers and regulate the duty cycles of the PWM signals.
With the above-mentioned hardware configurations and control setup,
this small wind generator is capable to provide artificial wind with both spatial and temporal variability.
As the goal of the current work is to investigate the aerodynamic characteristics of the artificially uniform flow generated from the fan array,
here we employ uniform duty cycles for all fans and perform flow measurements to evaluate the performance of the wind generator.

\subsection{Flow Measurement Techniques: Hotwire Anemometry and Particle Image Velocimetry} \label{sec:config}

\begin{figure}[!ht]
	\centering
	\begin{subfigure}{0.9\textwidth}
	\centering
		\includegraphics[height=140pt]{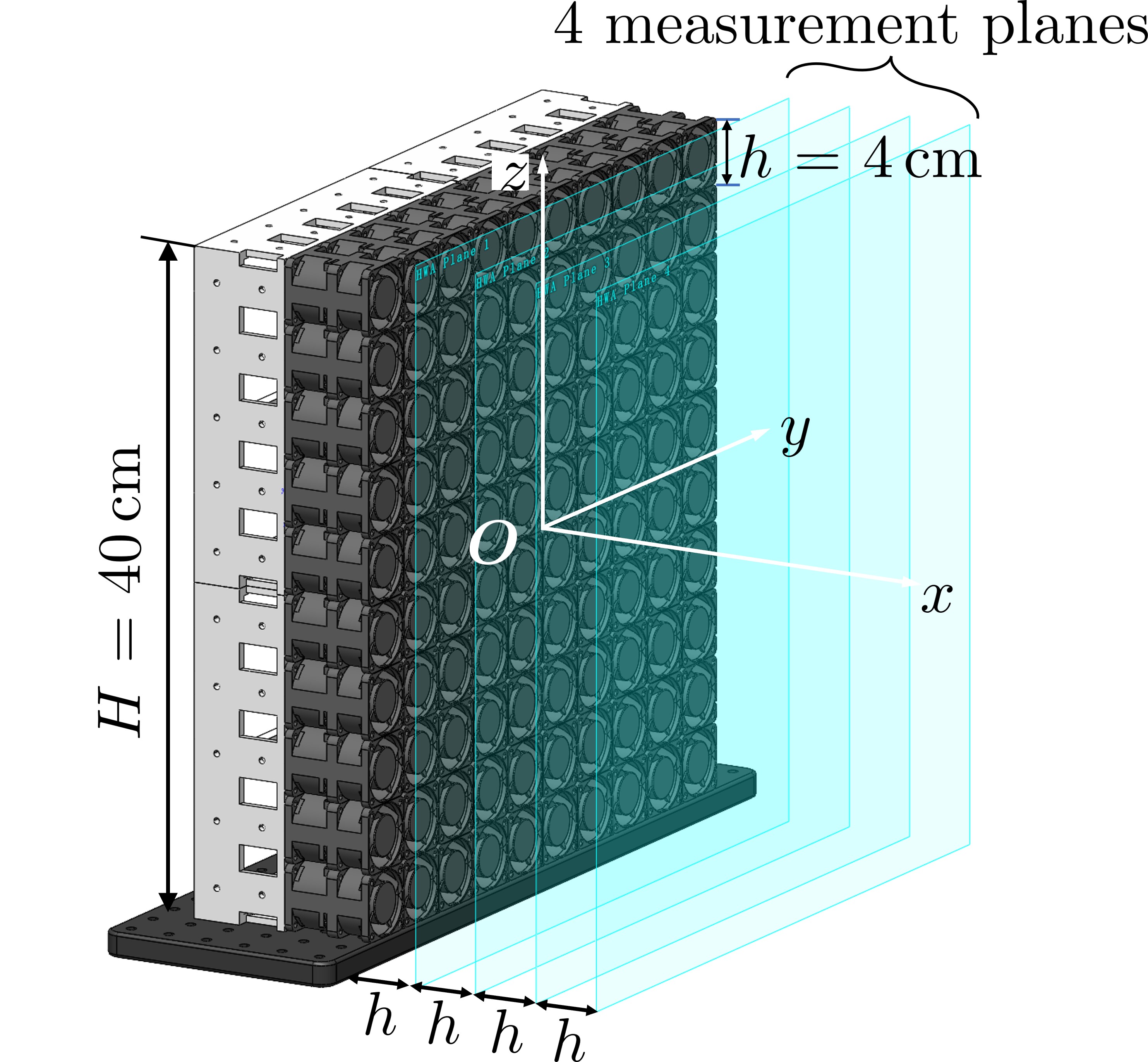}
		\caption{}
		\label{fig:HWA_Setup}
	\end{subfigure}
	\\
	\begin{subfigure}{0.9\textwidth}
	\centering
		\includegraphics[height=180pt]{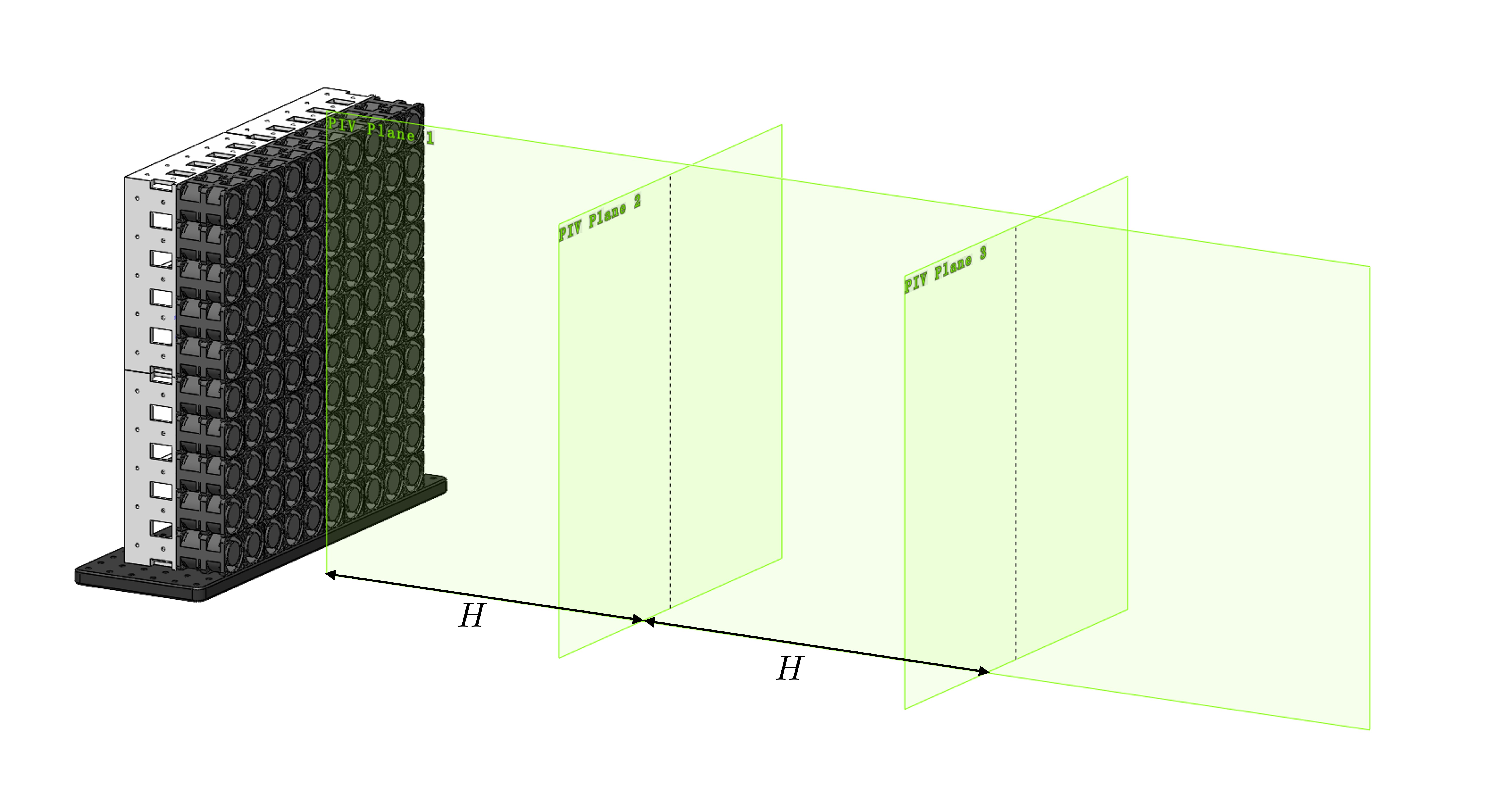}
		\caption{}
		\label{fig:PIV_Setup}
	\end{subfigure}

	\caption{A conclusion of flow measurement planes from HWA and PIV.
Here	$h$ denotes the height of fan elements and $H$ represents the height of the $10\times 10$ fan array.  
	(a) four HWA measurement planes in the near field; (b) three PIV measurement planes including one streamwise plane for planar PIV and two far field planes for stereoscopic PIV.
The Cartesian coordinate system used in this study is visualized in the figure.	}
	\label{fig:Exp_Setup}
\end{figure}

In this work, we utilize hotwire anemometry (HWA) and particle image velocimetry (PIV) to obtain accurate measurements of the flow fields.
An overview of all experimental campaigns performed in this study is presented in \Cref{tab:exp}.
To better describe the experimental setup, we introduce the Cartesian coordinate system originating from the center of fan array exit.
As displayed in \Cref{fig:HWA_Setup}, $x$ represents the streamwise direction, $y$ and $z$ represent horizontal and vertical directions, respectively.
We also adopt $h$ and $H$ to represent the height/width of the individual fan element and the entire fan array, respectively.
With the above-mentioned configurations, the Reynolds numbers ($Re_H$ ) are $3.07\times 10^5$,  $2.15\times 10^5$, and $1.11\times 10^5$ at $\SI{100}{\percent}$,  $\SI{80}{\percent}$, and  $\SI{50}{\percent}$ duty cycles, respectively.

Another important goal of this study is to investigate the flow characteristics in the near field of the flow.
As PIV measurements suffer from strong background reflection when the light sheet illuminates the cross-stream planes in the near field, hotwire measurements are adopted in this region.
We conduct hotwire measurements in conjunction with a two-dimensional positioning system to obtain streamwise velocity profiles near the fan exit. 
This approach produced promising flow field measurements with high spatial resolution.
A HangHua miniature single-sensor hotwire probe is used to measure the streamwise velocity signals.
The probe consists of a short tungsten wire with finite diameter and is attached to the two prongs made of stainless steel. 
As for the flow generated by fans, the streamwise fluctuations are an order of magnitude larger than the fluctuations in other directions \cite{Liang2023}. Thus, the results from the single hotwire measurement can be used to indicate the flow characteristics in the streamwise direction.
The measurements are performed on 4 near-field planes $x/h = 1$ to $4$.
To scan the flow field on each cross-stream plane, we move the hotwire probe with a spatial resolution of $dy = dz = \SI{4}{\milli\meter}$.
In this manner, the measurement results will form the velocity profiles on a two-dimensional grid.
To minimize the blockage effect introduced by the hotwire and the traverse system, we connect the hotwire sensor to a 15 cm long probe made of stainless steel, which is connected to a low-profile dual-rail traverse to ensure smooth motion in the flow field.
At each grid point, the sampling frequency is $F_s = \SI{4000}{\hertz}$ with a measurement time of $t=\SI{10}{\second}$.
In this study hotwire calibration is realized by logging the mean voltage output under different known incoming velocities, 
and a 5th order polynomial fit is performed to obtain the relationship between voltage outputs and the corresponding velocities.
The calibration is conducted using the wind generator, in which a Pitot tube and the hotwire probe measure the velocity and the voltage output at $(x,y,z)=(\SI{40}{\centi\meter},0,0)$, respectively.
In addition, the constant temperature anemometry is carefully tuned such that the hotwire is responsive to disturbances up to 2000 Hz from a square-wave test.
A National Instrument USB-6009 14-bit data acquisition device is used to sample voltage outputs from the hotwire anemometry.
The input range of the data acquisition card is $\pm \SI{10}{volt}$.
The overall measurement uncertainty from hotwire is about 3\% considering positioning error, calibration error, A/D resolution, and temperature variation ($\pm\SI{1}{\degreeCelsius}$).
Flow field measurements are conducted under a PWM duty cycle of $80\%$, as significant heat lumping will occur in the DC power at higher duty cycles during the extended scanning process.

In addition, particle image velocimetry (PIV) is employed to measure the velocity fields on one streamwise plane over the centerline ($y=0$), as well as two cross-stream planes in the far field of the fan array ($x/H=1,2$).
The measurement planes are visualized in \Cref{fig:PIV_Setup}.
In this study, we employ a LaVison PIV system equipped with a double-pulsed laser and two $2752 \times 2200$ mega-pixel CCD cameras.
For velocity measurements on the $y=0$ plane, a planar PIV configuration \citep{Raffel_1998} is adopted to obtain two-dimensional velocity measurements,
and two cameras are placed side-by-side to maximize the field of view (FoV) in the streamwise direction.
For far field measurements, a stereoscopic setup \citep{Raffel_1998} is used to calculate three dimensional velocity vectors. 
In this case, the two cameras with Scheimpflug adapters are mounted on opposite sides of the wind generator.
Calibration of the camera system is achieved by detecting and fitting the target points on a three-dimensional calibration plate a via a pinhole model.
 A dual-cavity Litron Nano L 200-15 Nd:YAG laser of $\SI{532}{\nano\meter}$ wavelength illuminates seeding particles from an Antari fog machine that produces particles about $\SI{0.2}{\micro\meter}$. 
PIV image pairs are acquired at a frequency of $\SI{12}{\hertz}$ under two representative duty cycles of $\SI{50}{\percent}$ and $\SI{100}{\percent}$.
At each duty cycle, a total of 3000 snapshots are acquired for the planar PIV measurements,  
while 2500 snapshots are recorded for stereoscopic PIV measurements on each far field plane.

All image pairs are post-processed in DaVis 10. 
PIV vectors are calculated using a multipass routine with 2 passes each for interrogation windows of $128 \times 128$ and $64 \times 64$. 
A $\SI{50}{\percent}$ overlap is also adopted during the post-processing, and the configurations results in a spatial resolution of $\SI{3.6}{\milli\meter}$ for planar and $\SI{6.5}{\milli\meter}$ for stereoscopic measurements.
The uncertainty analysis performed by DaVis is based on the cross-correlation statistics during the calculation of velocity vectors \citep{Wieneke_2015}. 
This analysis results in an uncertainty less than  $\SI{0.15}{\meter\per\second}$ for all PIV measurement campaigns.

\begin{table}[!ht]
  \centering
  \caption{An overview of experimental campaigns.}
  \begin{tabular}{ c c c c c }
    \hline \hline
    Experiment & Measurement Field(s) & Data Sampling & Spatial Resolution & Duty Cycle(s) \\
    \hline
    HWA & \begin{tabular}{@{}c@{}c@{}}$x/h=$ 1, 2, 3, 4, \\ $\lvert y/H \rvert \leqslant 0.5$, \\ $\lvert z/H \rvert \leqslant 0.5$ \end{tabular} & \begin{tabular}{@{}c@{}}$F_s = \SI{4000}{\hertz}$, \\ $t = \SI{10}{\second} $ \end{tabular} & \begin{tabular}{@{}c@{}}$\Delta y = \SI{5}{\milli\meter}$,  \\ $\Delta z = \SI{5}{\milli\meter}$  \end{tabular} & 80\%\\
\hline
    2D2C PIV & \begin{tabular}{@{}c@{}c@{}}$0.1\leqslant x/H\leqslant 3$, \\ $y/H = 0$, \\ $\lvert z/H \rvert \leqslant 0.75$ \end{tabular} & 3000 snapshots & \begin{tabular}{@{}c@{}}$\Delta x = \SI{3.6}{\milli\meter}$,  \\ $\Delta z = \SI{3.6}{\milli\meter}$  \end{tabular} & 50\%, 100\%\\
\hline
    2D3C PIV & \begin{tabular}{@{}c@{}c@{}}$x/H=$ 1, 2, \\ $\lvert y/H \rvert \leqslant 0.75$, \\ $\lvert z/H \rvert \leqslant 0.75$ \end{tabular} & 2500 snapshots & \begin{tabular}{@{}c@{}}$\Delta y = \SI{6.5}{\milli\meter}$,  \\ $\Delta z = \SI{6.5}{\milli\meter}$  \end{tabular} & 50\%, 100\%\\
    \hline \hline
\end{tabular}
\label{tab:exp}
\end{table}

\section{Results and Discussions}\label{sec:results}

In this section, we present the experimental results
 from PIV and HWA measurements,
and discuss the major characteristics of flow generated from SFAWG.
Experimental results and discussions about major flow characteristics are presented in \Cref{sec:statistics}.
The discussion focuses on the flow quality for the streamwise velocity component, as it is of most importance regarding the evaluation of artificial wind gusts.
The control of transverse velocity components 
for specific goals, like increased uniformity and maximized turbulence, 
represent one unique application niche of fan array wind generators.
The ongoing studies  will be subject of later publications.
Extended discussions regarding near field and far field turbulence properties are detailed in \Cref{sec:discussions}.

\subsection{Flow characterization}\label{sec:statistics}
\begin{figure}[!ht]
	\centering
	\begin{subfigure}{0.49\textwidth}
		\centering
		\includegraphics[width=.8\textwidth]{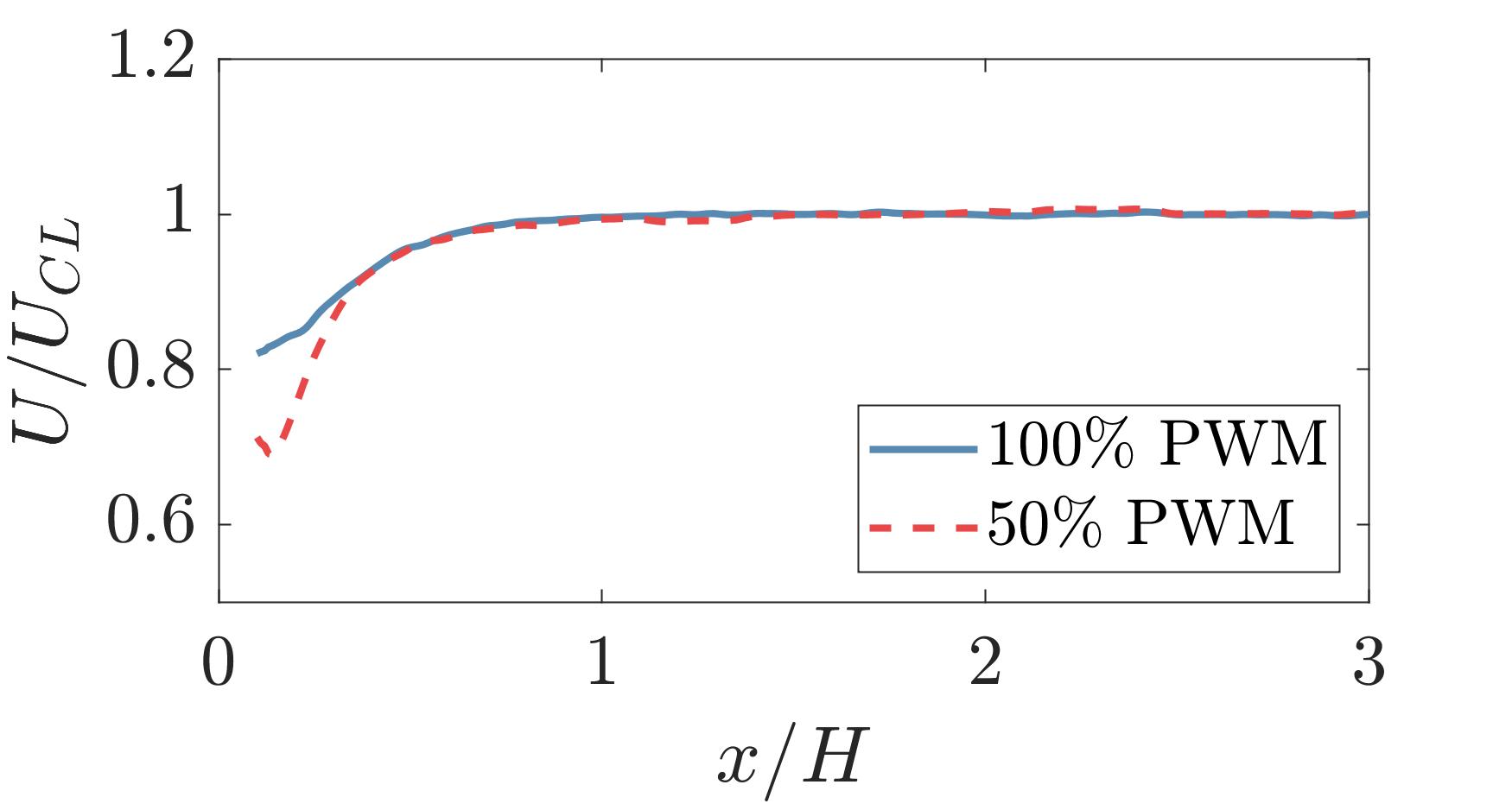}
		\caption{}
		\label{fig:CL_1}
	\end{subfigure}
	\begin{subfigure}{0.49\textwidth}
		\centering
		\includegraphics[width=.8\textwidth]{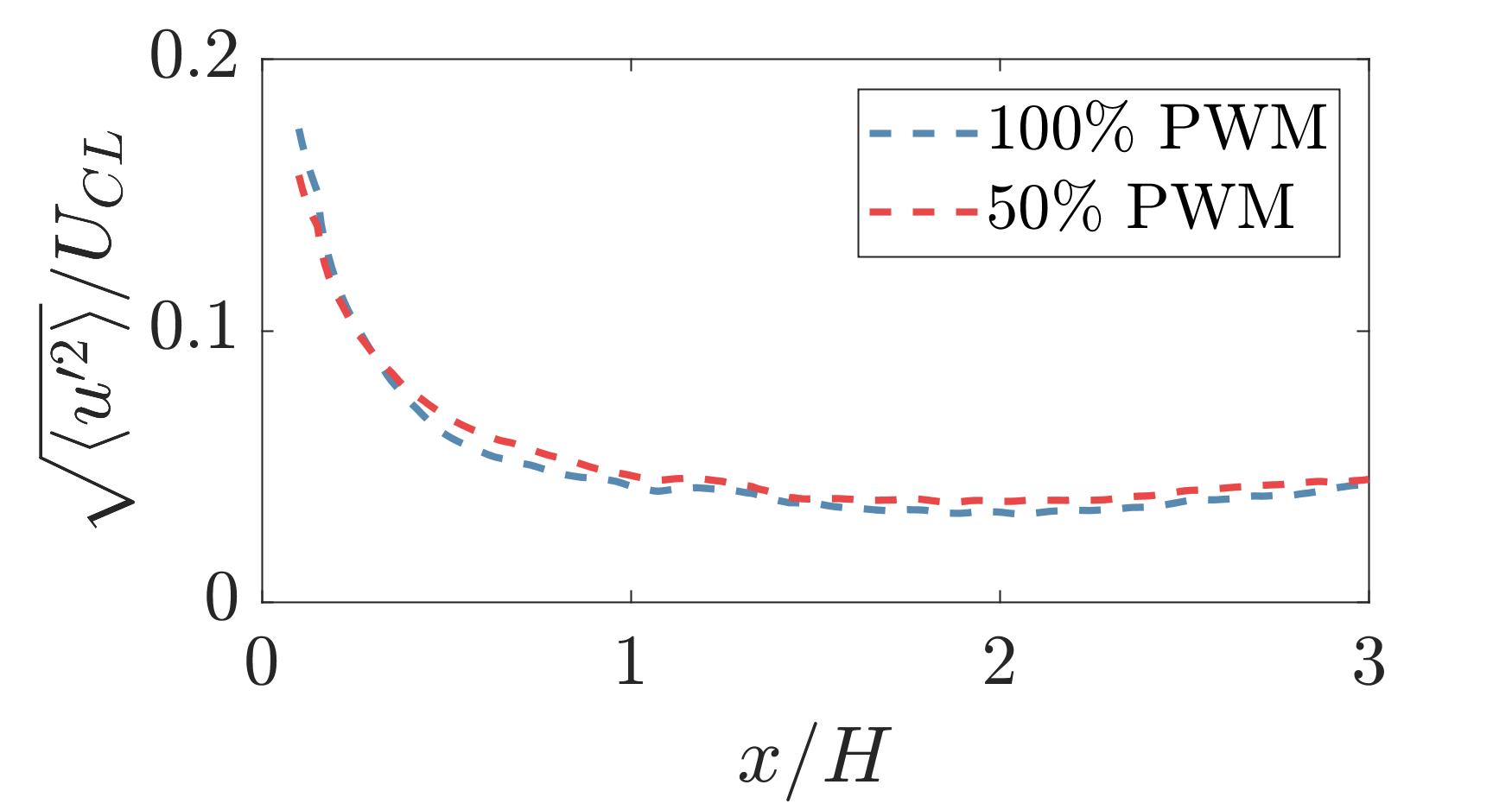}
		\caption{}
		\label{fig:CL_2}
	\end{subfigure}

	\caption{(a) Centerline streamwise velocity normalized by the far field centerline velocity $U_{CL}$, and (b) centerline mean-squared turbulent velocity $\sqrt{\langle u'^2 \rangle}$ normalized by $U_{CL}$.}
	\label{fig:CL}
\end{figure}

From PIV snapshots that are measured according to the configuration presented in \Cref{tab:exp}, the mean streamwise velocity $U$, as well as the fluctuating counterpart $u'$, can be calculated based on the Reynolds decomposition \citep{Pope_2000}.
\Cref{fig:CL_1} and \Cref{fig:CL_2} present the evolution of streamwise velocity ($U$), and mean-squared turbulent velocity ($\sqrt{\langle u'^2 \rangle}$) along the centerline of the fan array ($y=0$, $z=0$), respectively.
Here $\langle \cdot \rangle$ represents the ensemble average from PIV snapshots.
At the same time, we define and calculate the far field jet centerline velocity $U_{CL}$ as the spatially averaged centerline velocity between $x=1H$ and $3H$,
and we normalize both $U$ and $\sqrt{\langle u'^2 \rangle}$  by the far field centerline velocity in both \Cref{fig:CL_1} and \Cref{fig:CL_2}.
Based on our calculation, $U_{CL}$ is about $\SI{12.24}{\meter\per\second}$ at 100\% duty cycle, and is about $\SI{4.43}{\meter\per\second}$ at 50\% duty cycle.
After normalization, the velocity profiles at 50\% and 100\% duty cycles present similar characteristics.
Before $x/H=1$, a gradual increment of the jet centerline velocity is observed which corresponds to the initial mixing stage in the near field ($x=\mathcal{O}(h)$).
In this region, small annular jets with high velocity are generated from the fan ducts. 
These jets expand and gradually increase the velocity in the rest of the flow region, including the centerline.
The mixing process becomes weaker as the small jets propagate downstream, and it is accompanied by the decrement of mean-squared turbulence velocity on the centerline.
The mixing process appears to stop near $x=1H$, where the streamwise velocity reaches $U_{CL}$, and the mean-squared turbulent velocity drops to about 3.5\% of the far field centerline velocity.
Thereafter, both $U$ and $\sqrt{\langle u'^2 \rangle}$ remain nearly constant between $1H$ and $3H$.
The plateau of both the centerline velocity and the turbulence velocity between $1H$ and $2H$ is similar to the typical characteristics of a turbulent jet few diameters downstream (see, for example, \cite{Proena_2019}).
This similarity will be elaborated in the following.

\begin{figure}[!htb]
	\centering
	\includegraphics[width = .75\textwidth]{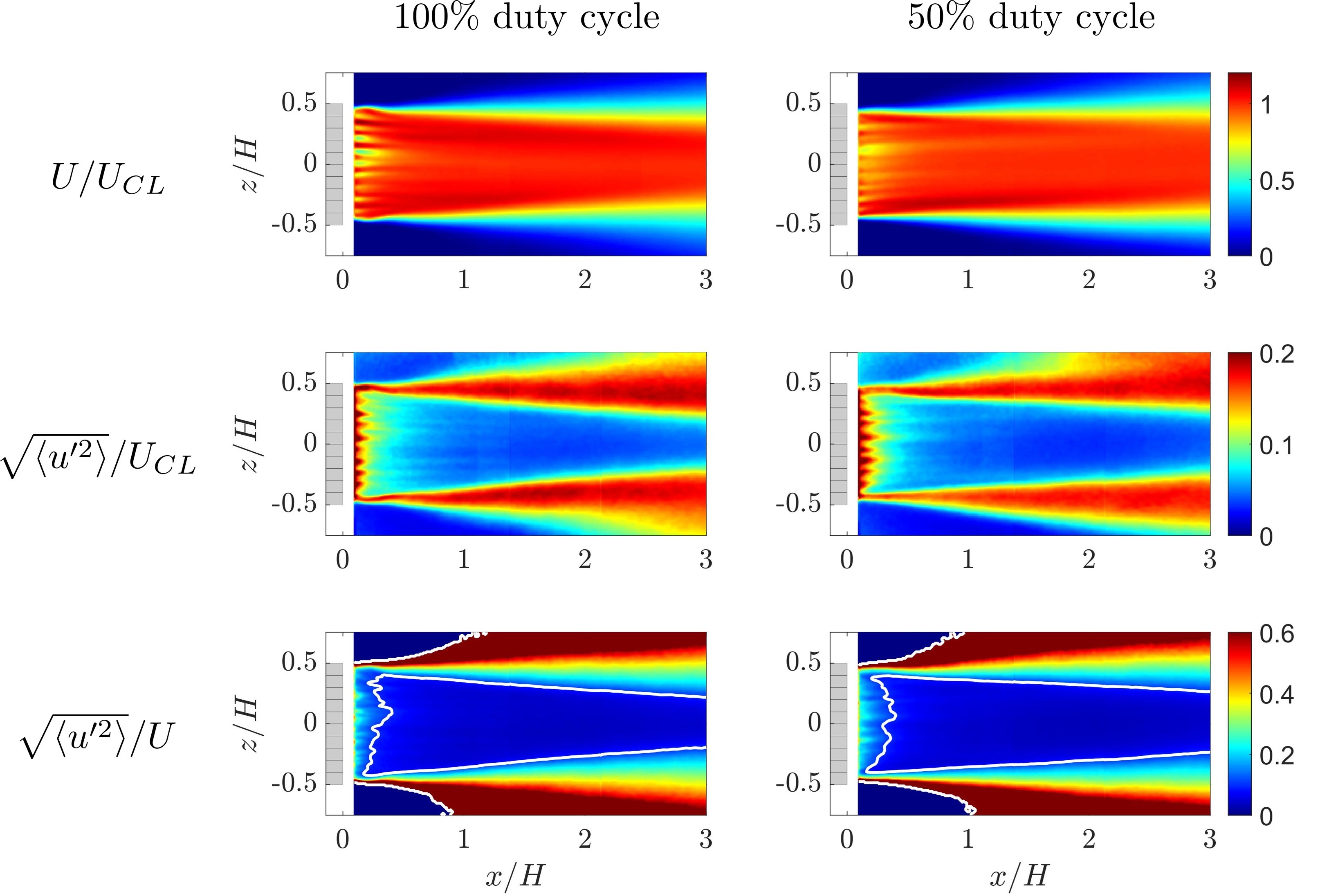}
	\caption{A comparison of flow properties on the streamwise PIV measurement plane at duty cycles of 100\% and 50\%. First row: normalized mean streamwise velocity;
	second row: normalized mean-squared turbulent velocity; third row: turbulence intensity. White isolines on the third row represents a turbulence intensity of 10\%. Fan locations are visualized as gray rectangles in the figure.}
	\label{fig:2D2C}
\end{figure}

\Cref{fig:2D2C} displays the mean streamwise velocity (first row), mean-squared turbulent velocity (second row), and turbulence intensity (third row) that are measured under 100\% and 50\% duty cycles via planar PIV.
The flow contours under both duty cycles exhibit strong similarities, which include the initial mixing in the near field, as well as the expansion of top and bottom shear layers in the far field ($x= \mathcal{O}(H)$).
The nearly symmetrical expansion of the shear layers, as well as the streamwise decay of the ''potential core'', possesses similar characteristics as a typical jet flow.
However, the mixing process in the near field are not completely symmetrical about the centerline, and local non-uniformity can also be observed inside the core region.  
These asymmetries may originate from slight differences of the fan elements.
Similar to a typical jet flow, the core region  possesses relatively low turbulence level in the far field of the fan array.
In contrast, the shear layers are of high turbulence levels, where complex, multi-scaled turbulent structures emerge, evolve, and interact with each other.
As the goal of the current work is to generate uniform flow from the fan array, flow regions with uniform velocity profiles and low turbulence intensity levels are preferred.
To visualize the region of low turbulence intensity, we use white isolines in the last row of \Cref{fig:2D2C} to represent a turbulence level of 10\%. 
As the flow progresses downstream, the vertical extent of the region of low turbulence becomes increasingly restricted.
At $x=3H$, the vertical extent of the low turbulence region is only approximately $0.5H$ for both duty cycles, 
indicating that the effective cross-sectional area with low turbulence level is reduced to only 25\% of the fan array's cross-sectional area.

\begin{figure}[!htb]
	\centering
	\includegraphics[width = .99\textwidth]{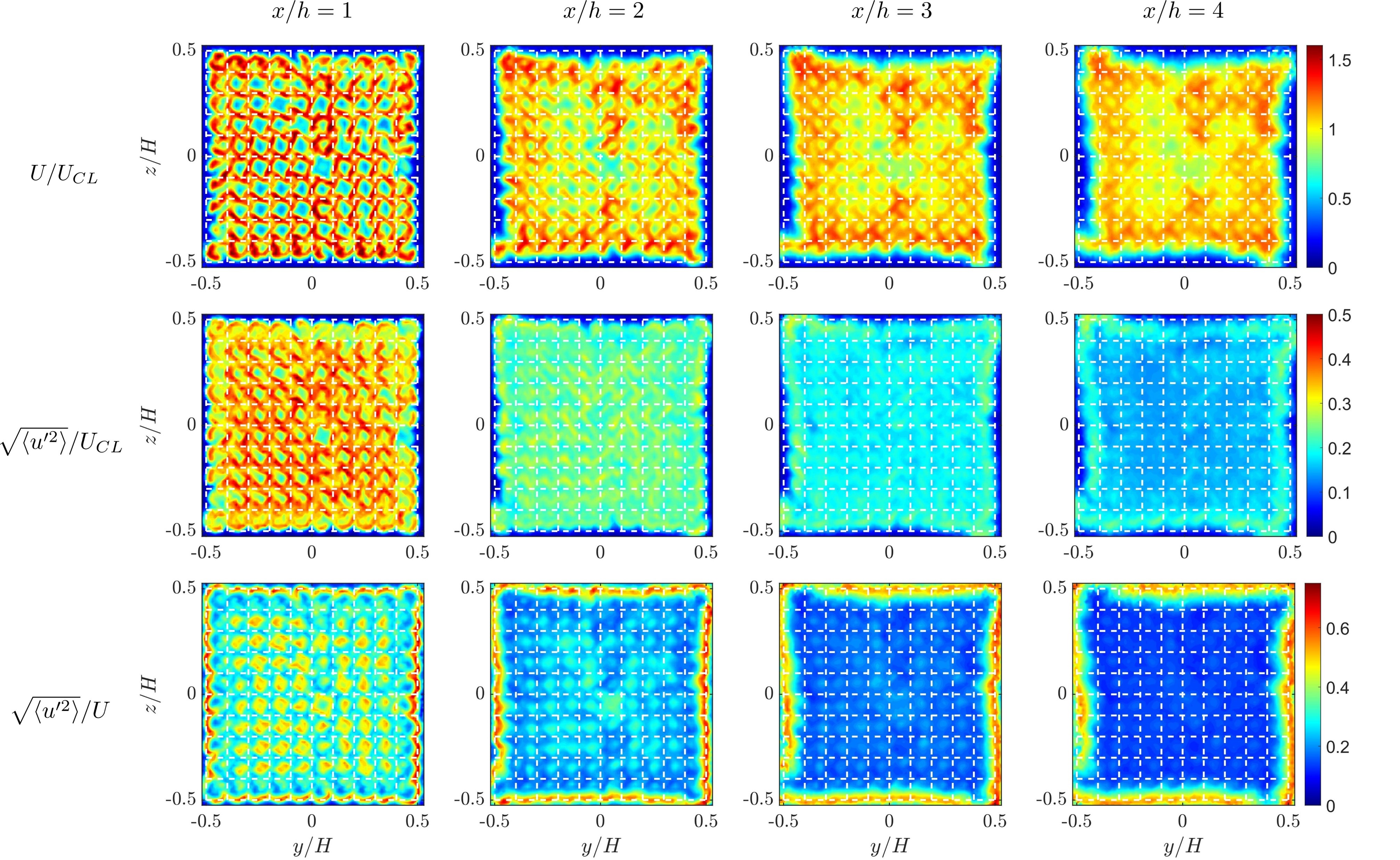}
	\caption{Flow profiles at four near field planes $x/h = 1,2,3,4$ from hotwire measurements at 80\% duty cycle. First row: normalized mean streamwise velocity;
			second row: normalized mean-squared turbulent velocity; third row: turbulence intensity. Fan locations are visualized by the white grid lines.}
	\label{fig:HWA}
\end{figure}

To investigate the initial mixing in the near field, hotwire measurement results from four near field planes at $x/h=1,2,3,4$ are presented in \Cref{fig:HWA}. 
At a downstream distance of $1h$, the flow is strongly non-uniform, with high-speed regions mostly concentrated around the annular fan duct. 
Due to subtle differences among fan elements, the flow profiles generated by different fans exhibit slight variations. 
As the flow progresses in the streamwise direction, the high-speed flow gradually interacts and mixes with the low-speed fluid, accompanied by a decrease in turbulence level from $1h$ to $4h$ downstream. 
Another interesting observation is the distortion of the flow envelope as the flow moves downstream. 
At $x=1h$, the spatial locations of the annular jets are consistent with the location of the fan array, as visualized by the white grid lines. 
However, with increased mixing, the nearly-rectangular overall profile becomes more distorted at downstream locations, with the edges shrinking inward and the corners stretching outward. 
This is caused by the swirling of the flow generated by the rotating fan blades. 
Similar observations are also found in the turbulence intensity profiles, with regions of high turbulence intensity corresponding to the appearance of the turbulent shear layer.

\begin{figure}[!htb]
	\centering
	\includegraphics[width = .99\textwidth]{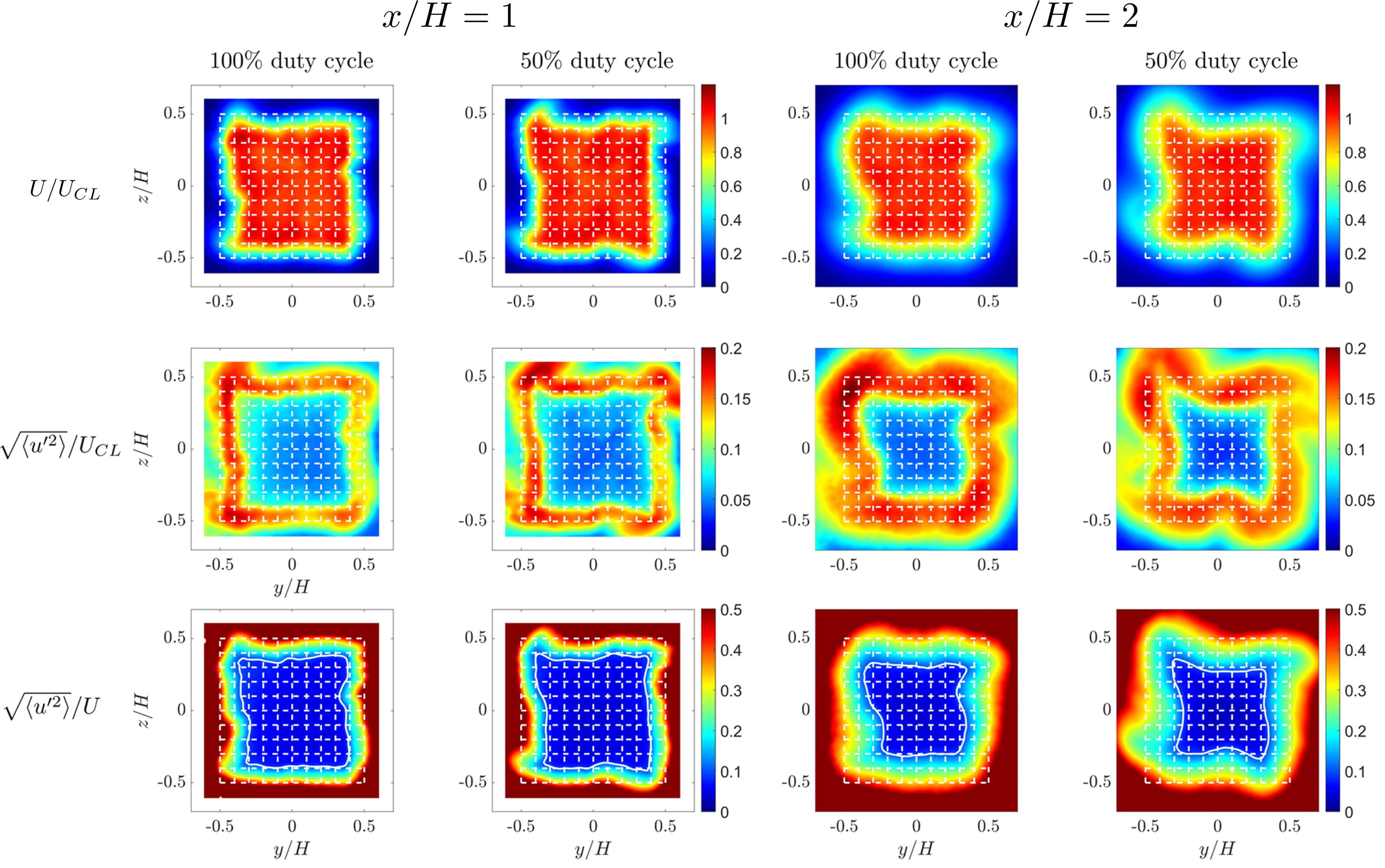}
	\caption{Flow profiles at far field planes $x/H = 1,2$ from stereoscopic PIV measurements at 50\% and 100\% duty cycles. First row: normalized mean streamwise velocity;
		second row: normalized mean-squared turbulent velocity; third row: turbulence intensity. White isolines on the third row represents a turbulence intensity of 10\%. Fan locations are visualized by the white grid lines.
	}
	\label{fig:2D3C}
\end{figure}

As the flow progresses to the far field, the cross-stream flow profiles become more and more distorted. In \Cref{fig:2D3C}, the mean streamwise velocity, mean-squared turbulent velocity, and turbulence intensity measured on two far field planes at $x/H=1,2$ via stereoscopic PIV are displayed. 
After the mixing process in the near field, the core region in the far field becomes more uniform and less turbulent. 
Meanwhile, the expansion and distortion of the annular shear layer dominate the far field flow characteristics. 
The asymmetric profile may originate from the swirling of the flow generated from the small fans. Attenuating this distortion of the flow will become an important topic in order to produce wind profiles with higher quality.
From $1H$ to $2H$ downstream, the velocity profile continues to be stretched and distorted, while the significant thickening of the shear layer can be observed under both duty cycles. In addition, the region of low turbulence intensity inside the core region shrinks rapidly, which is similar to the observations in \Cref{fig:2D2C}. 
We calculate the vertical span of the low turbulence intensity region at different streamwise locations based on the white isolines which represents a turbulence intensity of 10\%.
Under both duty cycles, the results translate to a cross-sectional area of $0.36H^2$ at $x=2H$ and $0.25H^2$ at $x=3H$ .

\subsection{Extended discussions on turbulence characteristics}\label{sec:discussions}
\begin{figure}[!ht]
	\centering
	\begin{subfigure}[c]{0.25\textwidth}
		\centering
		\includegraphics[width=.99\textwidth]{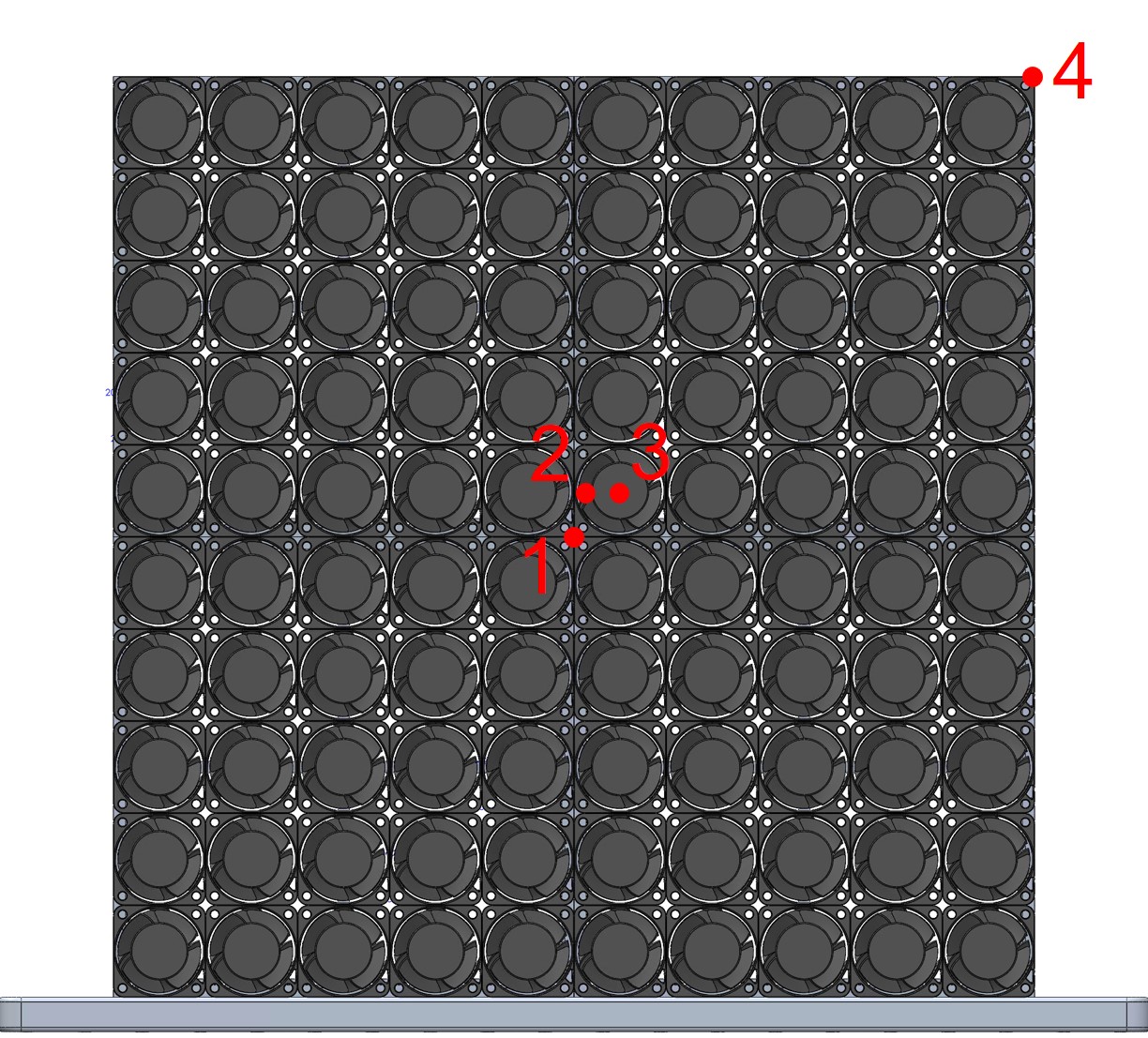}
		\caption{}
		\label{fig:PSD_1}
	\end{subfigure}
	\begin{subfigure}[c]{0.74\textwidth}
		\centering
		\includegraphics[width=.99\textwidth]{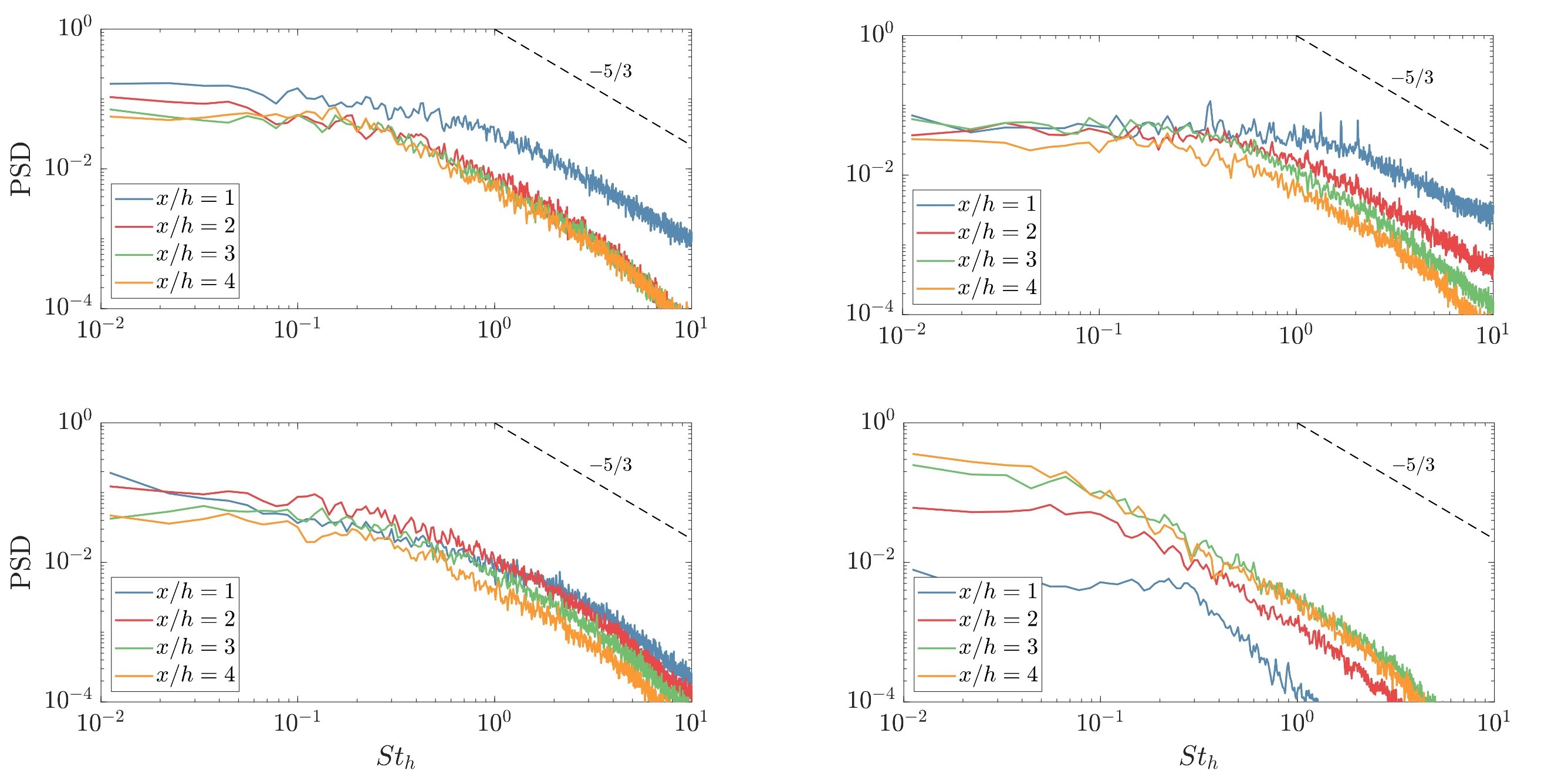}
		\caption{}
		\label{fig:PSD_2}
	\end{subfigure}

	\caption{(a) The four points used to calculate the power spectral density (PSD) from hotwire measurements. The $(y,z)$ coordinates of the four points are: $(0,0)$,$(\SI{0.5}{\centi\meter},\SI{2}{\centi\meter})$,$(\SI{2}{\centi\meter},\SI{2}{\centi\meter})$, and $(\SI{20}{\centi\meter},\SI{20}{\centi\meter})$.(b) Power spectal density of points 1 to 4 at different near field locations.}
	\label{fig:PSD}
\end{figure}

In this subsection we continue the discussion about  turbulent characteristics in the near and far fields of the wind generator.
Based on the observed strong similarity across different duty cycles, we will only present results from the maximum duty cycles (80\% for hotwire and 100\% for PIV measurements) to demonstrate the turbulence characteristics.
Using hotwire measurements, we analyze the power spectrum density (PSD) of the velocity signals at four representative points on the $(y,z)$ plane: (1) the center of the cross-stream plane at $(0,0)$, (2) the center of the annular duct at $(\SI{0.5}{\centi\meter},\SI{2}{\centi\meter})$, (3) the center of a fan element at $(\SI{2}{\centi\meter},\SI{2}{\centi\meter})$, and (4) the corner of the fan array at $(\SI{20}{\centi\meter},\SI{20}{\centi\meter})$, as illustrated in  \Cref{fig:PSD_1}. The PSDs of these four points at different near field planes are presented in \Cref{fig:PSD_2}. 
The calculation of PSDs follows the typical windowing and averaging procedure.
The Hanning window is adopted in the PSD calculation, with an overlap of 50\%.
The frequency $f$ is non-dimensionalized forming the Strhouhal number $St_h =fh/U_{CL}$, and PSD is also non-dimensionalized according to these scaling parameters.
The use of the $St_h$ is grounded on the similarity between the typical jet flow and the small "jets" generated by the individual fans.
In addition, we include Kolmogorov's $-5/3$ power law as black dashed lines in the figure.
For nearly all cases, flat energy distributions occur in the low-frequency region, followed by exponential rolling-offs starting around $St_h=0.1$.
This implies that flow generated by the fay array is governed by complex turbulent structures at a wide range of turbulence scales.
Notably, for point 4, we observe a broadband peak near $St_h=0.3$ at $x/h=1$. 
This peak may be related to the dynamics of the small annular jet, according to the jet characteristic frequency documented in \citep{Hussain_1981}. 
However, the peak is quickly overwhelmed by broadband turbulence as the flow moves further downstream.
At $4h$ downstream, the roll-offs of the power spectra follow the$-5/3$ power law, which indicates the end of the local mixing phenomenon among small "jets" generated by the individual fans.

\begin{figure}[!htb]
	\centering
	\includegraphics[width = .99\textwidth]{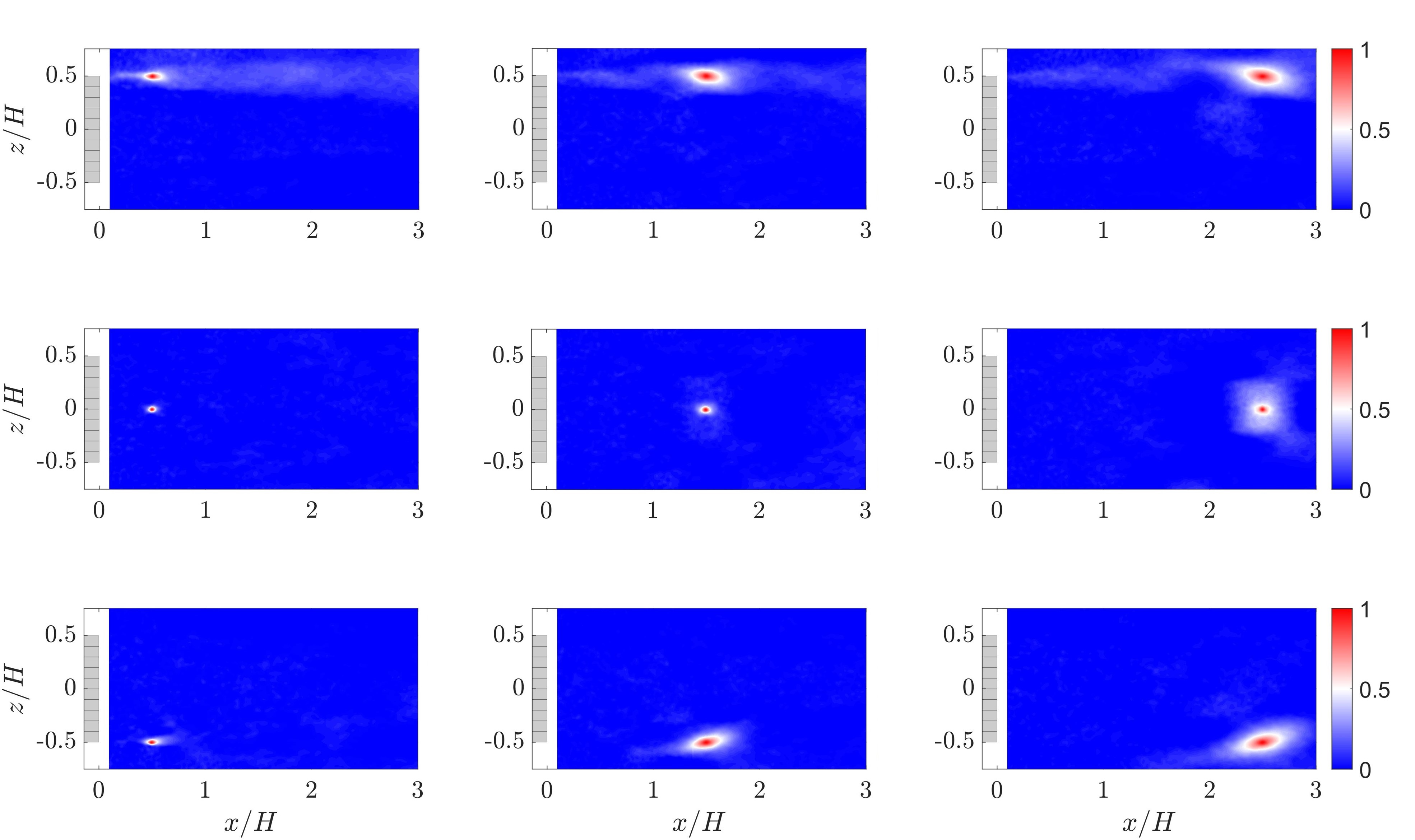}
	\caption{Normalized streamwise velocity correlation coefficient in the $y=0$ plane at 9 selected $(x,z)$ points under 100\% duty cycle.}
	\label{fig:xcorr}
\end{figure}

To reveal the spatial length scale of turbulent structures, the PIV snapshots measured on the streamwise plane is applied to calculate the normalized streamwise velocity correlation coefficients at 9 representative points.
The origins of these 9 points are fixed at three streamwise locations ($x/H=0.5,1.5,2.5$ ) and three vertical locations ($z/H=0.5,0,-0.5$).
These points will allow us to examine the streamwise evolution of the velocity correlation on the upper shear layer, centerline, and the lower shear layer.
The normalized velocity correlation coefficient ($R$) is calculated based on the following equation:
\begin{equation}	
	R(\mathbf{x},\mathbf{x_0}) = \frac{\langle u'(\mathbf{x},t_n)u'(\mathbf{x_0},t_n) \rangle}{\sqrt{\langle u'(\mathbf{x},t_n)^2 \rangle} \sqrt{\langle u'(\mathbf{x_0},t_n)^2 \rangle}}.
\end{equation}
Here $\mathbf{x}$ represents the locations of all possible points on the measurement plane and $\mathbf{x_0}$ represents the locations of the 9 correlation points.
These correlations are displayed in \Cref{fig:xcorr}.
On the flow centerline, the spatial extension of the streamwise correlation remains small along the streamwise direction.
However, clear growths of the correlation region can be observed in upper and lower shear layers.
In addition, the streamwise correlations in the shear layers  incline at an angle around $\SI{15}{\degree}$ about the jet axis.
These observations closely resemble the behavior of a turbulent jet, as noted by \citep{Ukeiley_2007}.
Therefore, we can conclude that the shear layer is subject to jet dynamics, even though it is initially distorted and stretched due to swirling.

\begin{figure}[!ht]
	\centering
	\begin{subfigure}[c]{0.4\textwidth}
		\centering
		\includegraphics[width=.99\textwidth]{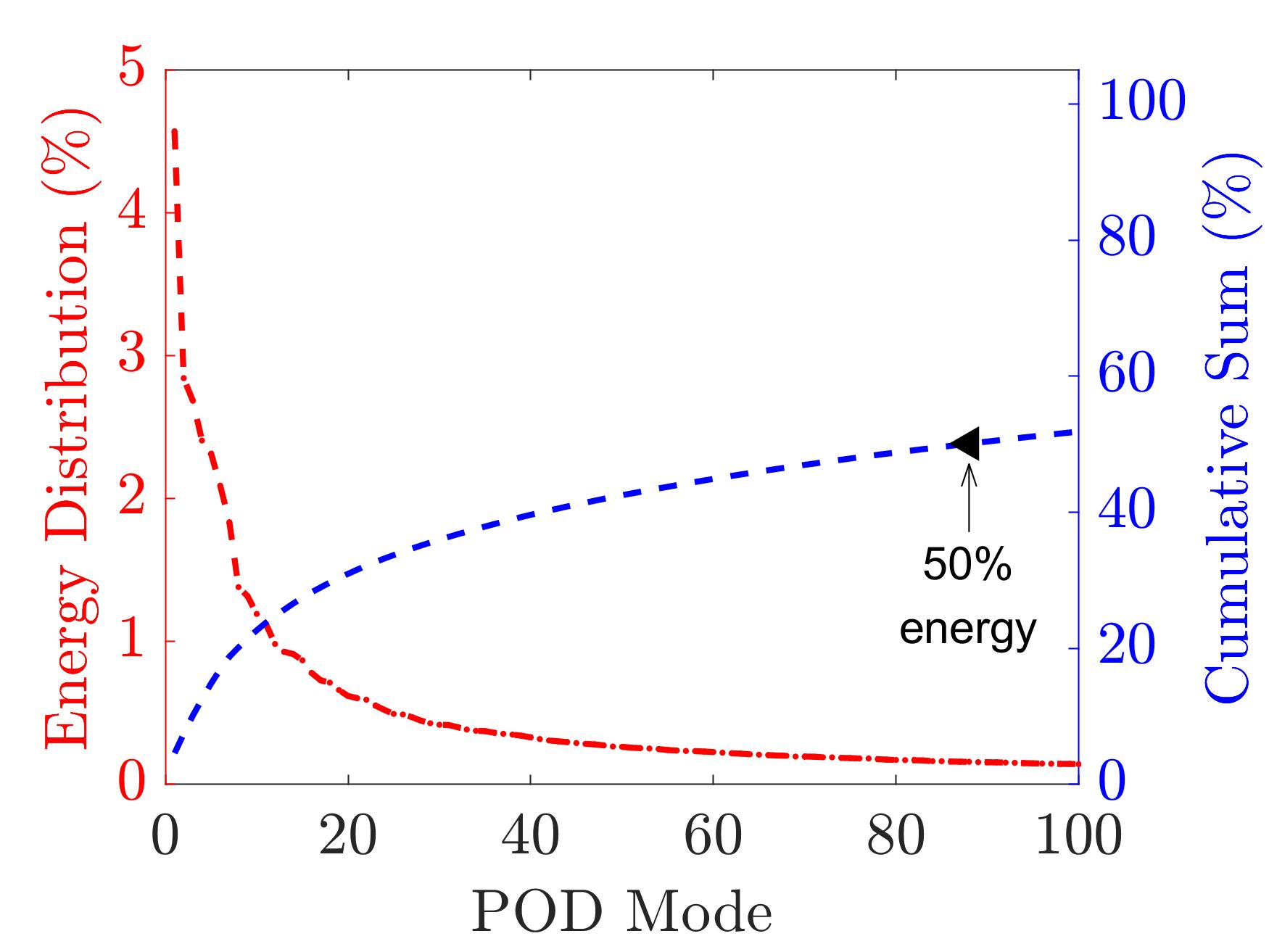}
		\caption{}
		\label{fig:POD_energy}
	\end{subfigure}\\
	\begin{subfigure}[c]{0.95\textwidth}
		\centering
		\includegraphics[width=.99\textwidth]{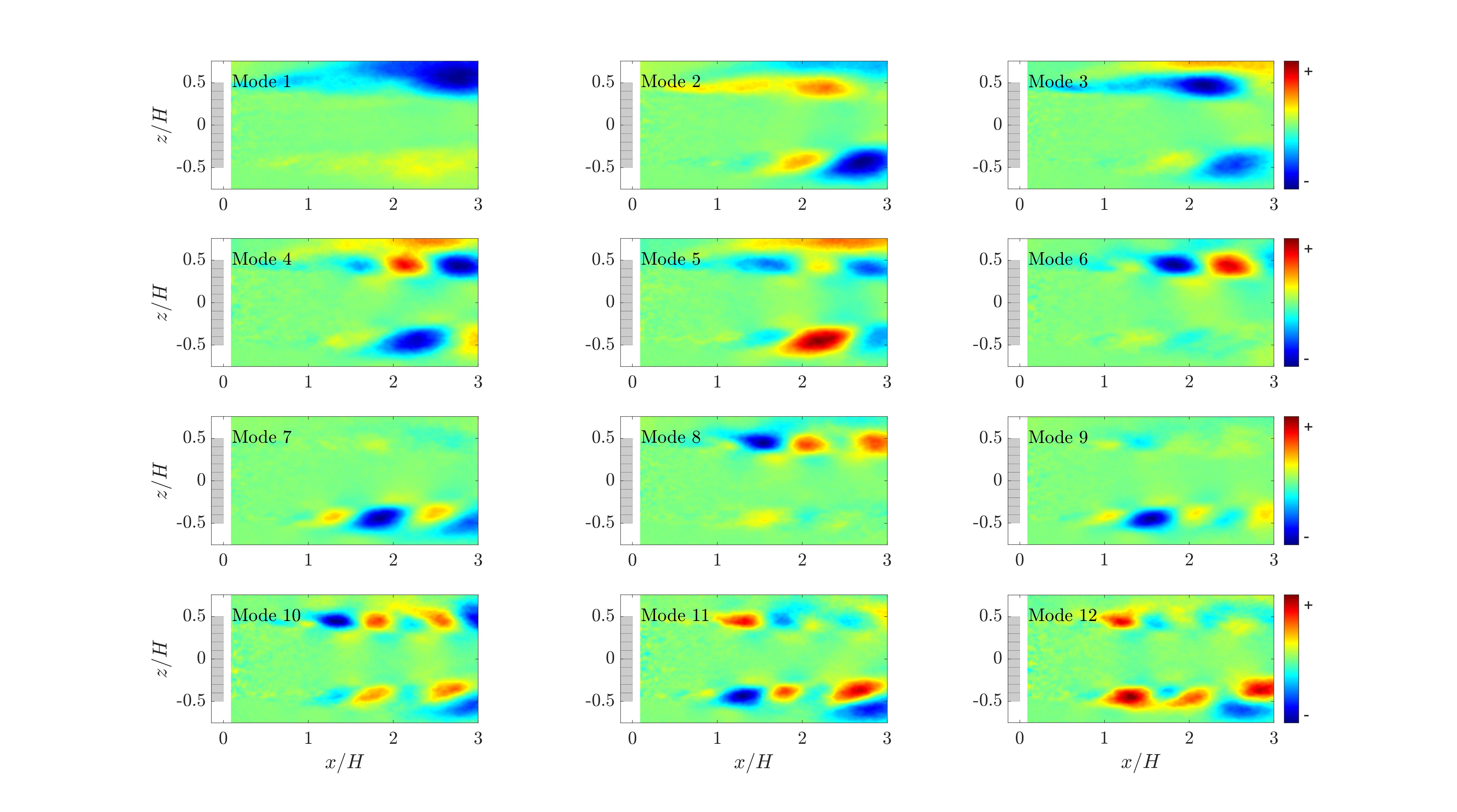}
		\caption{}
		\label{fig:POD_modes}
	\end{subfigure}

	\caption{Results of the snapshot POD from PIV measurements on the streamwise plane at 100\% PWM. (a) energy distribution of the leading POD modes; (b) streamwise components of the first 12 POD modes.}
	\label{fig:POD}
\end{figure}

In addition, we perform the snapshot proper orthogonal decomposition (POD, \citep{Lumley_1967,Sirovich_1987})  to extract the dominant coherent structures from the PIV images recorded on the streamwise plane at 100\% duty cycle.
The snapshot POD is a data analysis technique to decompose the velocity vectors into a set of orthonormal spatial modes $\mathbf{u}_i$ and the corresponding mode amplitudes $a_i$, such that:
\begin{equation}
	\mathbf{u'}(\mathbf{x},t_n) \approx \sum_{i=1}^{N}a_i(t_n) \> \mathbf{u}_i(\mathbf{x}) .
\end{equation}
Here $\mathbf{u'}$ represents the turbulent velocity vector 
and $t_n$ represents the time for the $n$th snapshot. 
$N$ is the truncation of POD modes.
We note that for the calculation of POD modes, the full velocity vectors from PIV measurements are employed to calculate the POD modes, and in the following discussions we only present the eigenvector in the streamwise direction which can already reveal the shape and distribution of turbulent structures in the flow.
We visualize the energy distribution of the leading POD modes in \Cref{fig:POD_energy} and display the streamwise components of the first 12 POD modes in \Cref{fig:POD_modes}. The energy distribution of the leading POD modes, as shown in \Cref{fig:POD_energy}, is highly dispersive.
Specifically, the first two POD modes contains only 4.5\% and 2.8\% of the overall turbulent kinetic energy (TKE), respectively. 
The need for the first 88 modes to account for 50\% of the TKE suggests the existence of turbulent structures with multiple scales and divergent dynamics, which is consistent with observations from hotwire measurements.
Additionally, \Cref{fig:POD_modes} shows the dominant flow structures with the highest energy levels. In the leading POD modes, turbulent structures are mainly located inside the top and bottom shear layers, and their spatial scales become smaller for higher-order modes. However, none of the spatial modes display a clear symmetrical or anti-symmetrical pattern, which might be caused by the distortion of the shear layer profiles depicted in \Cref{fig:2D3C}.

\begin{figure}[!ht]
	\centering
	\begin{subfigure}[c]{0.34\textwidth}
		\centering
		\includegraphics[width=.99\textwidth]{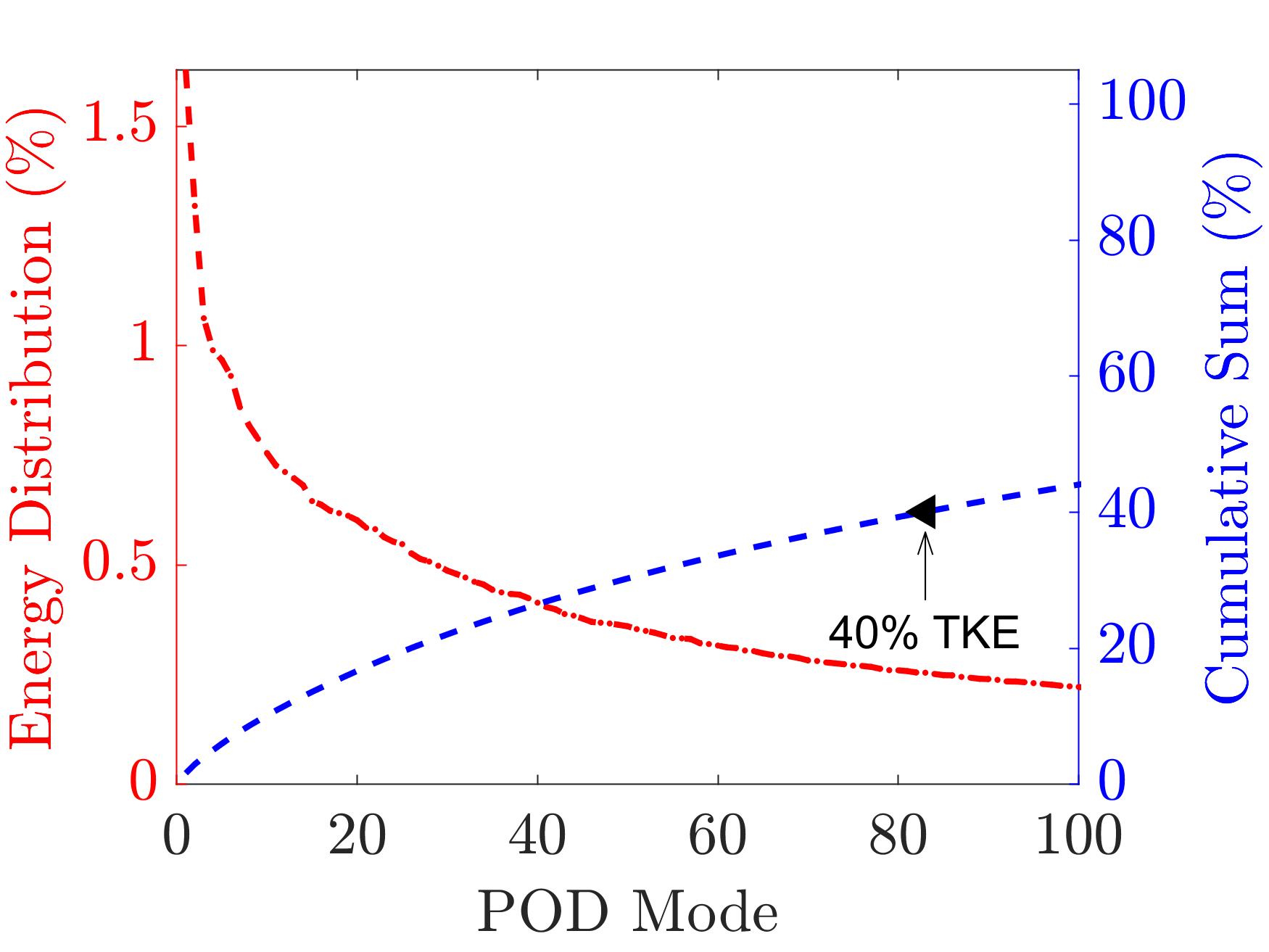}
		\caption{}
		\label{fig:POD_energy1H}
	\end{subfigure}
	\begin{subfigure}[c]{0.64\textwidth}
		\centering
		\includegraphics[width=.99\textwidth]{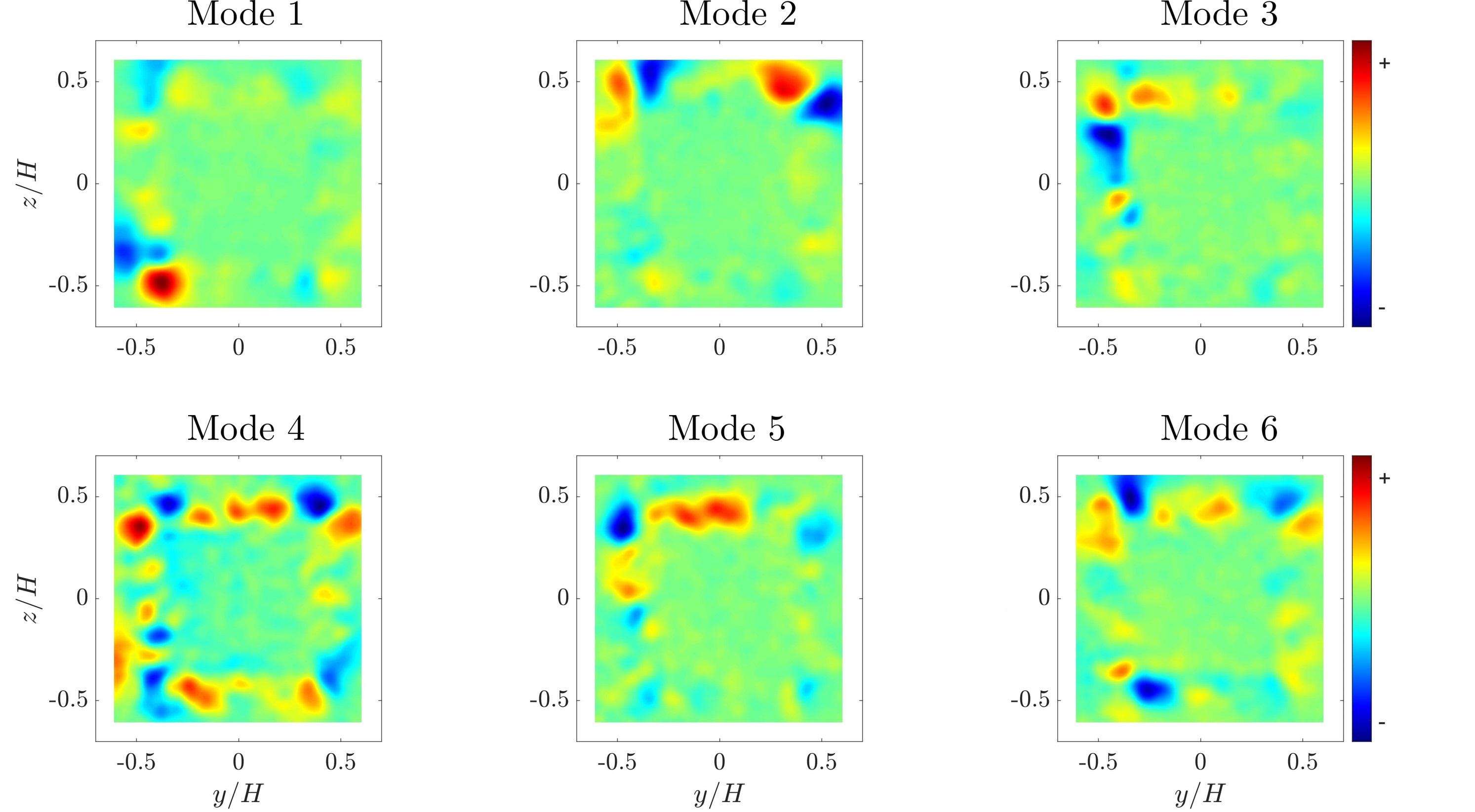}
		\caption{}
		\label{fig:POD_modes1H}
	\end{subfigure}

	\caption{Results of the snapshot POD from PIV measurements on the cross-stream plane $x=1H$ at 100\% PWM. (a) energy distribution of the leading POD modes; (b) streamwise components of the first 6 POD modes.}
	\label{fig:POD1H}
\end{figure}

\begin{figure}[!ht]
	\centering
	\begin{subfigure}[c]{0.34\textwidth}
		\centering
		\includegraphics[width=.99\textwidth]{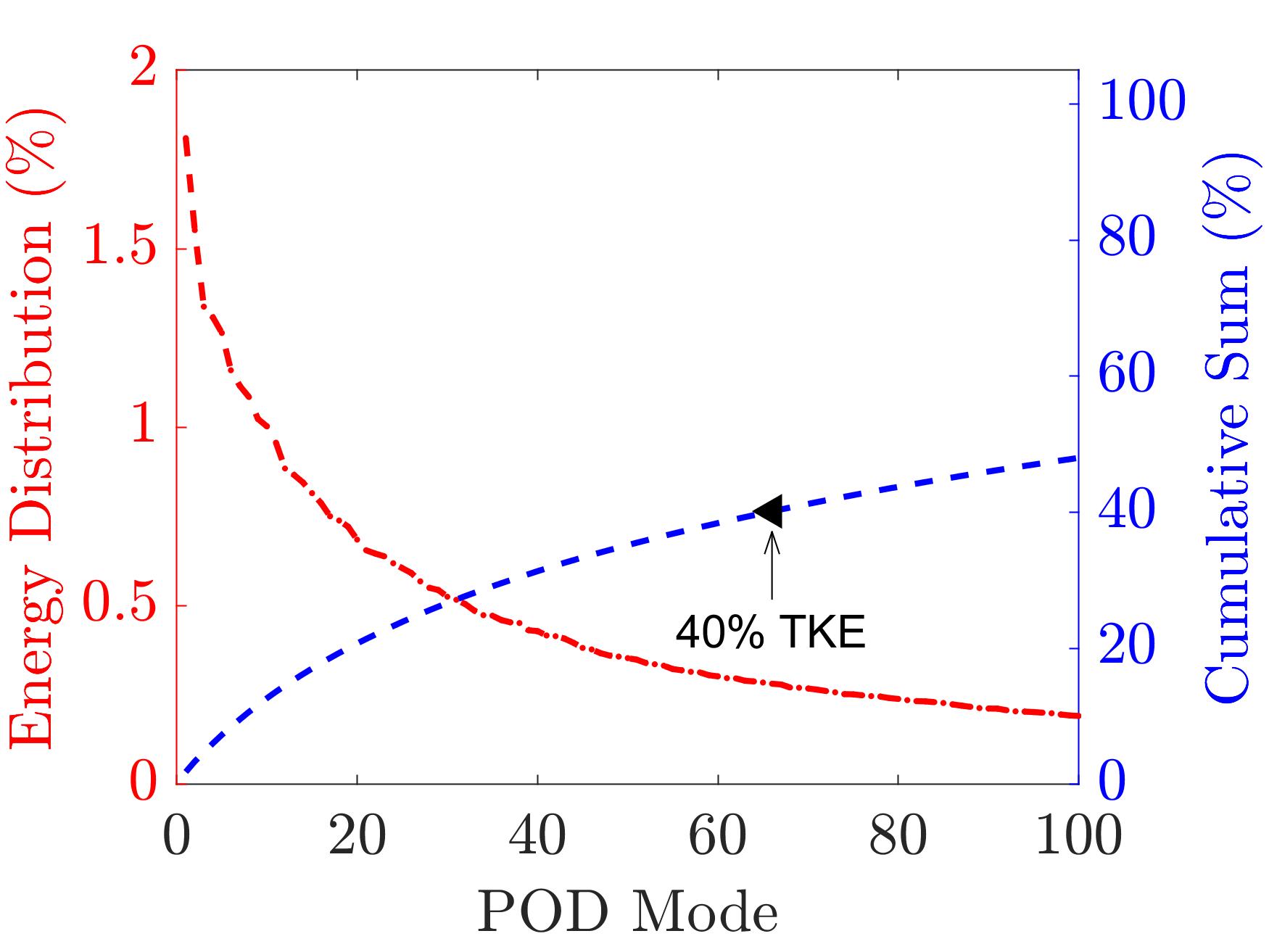}
		\caption{}
		\label{fig:POD_energy2H}
	\end{subfigure}
	\begin{subfigure}[c]{0.64\textwidth}
		\centering
		\includegraphics[width=.99\textwidth]{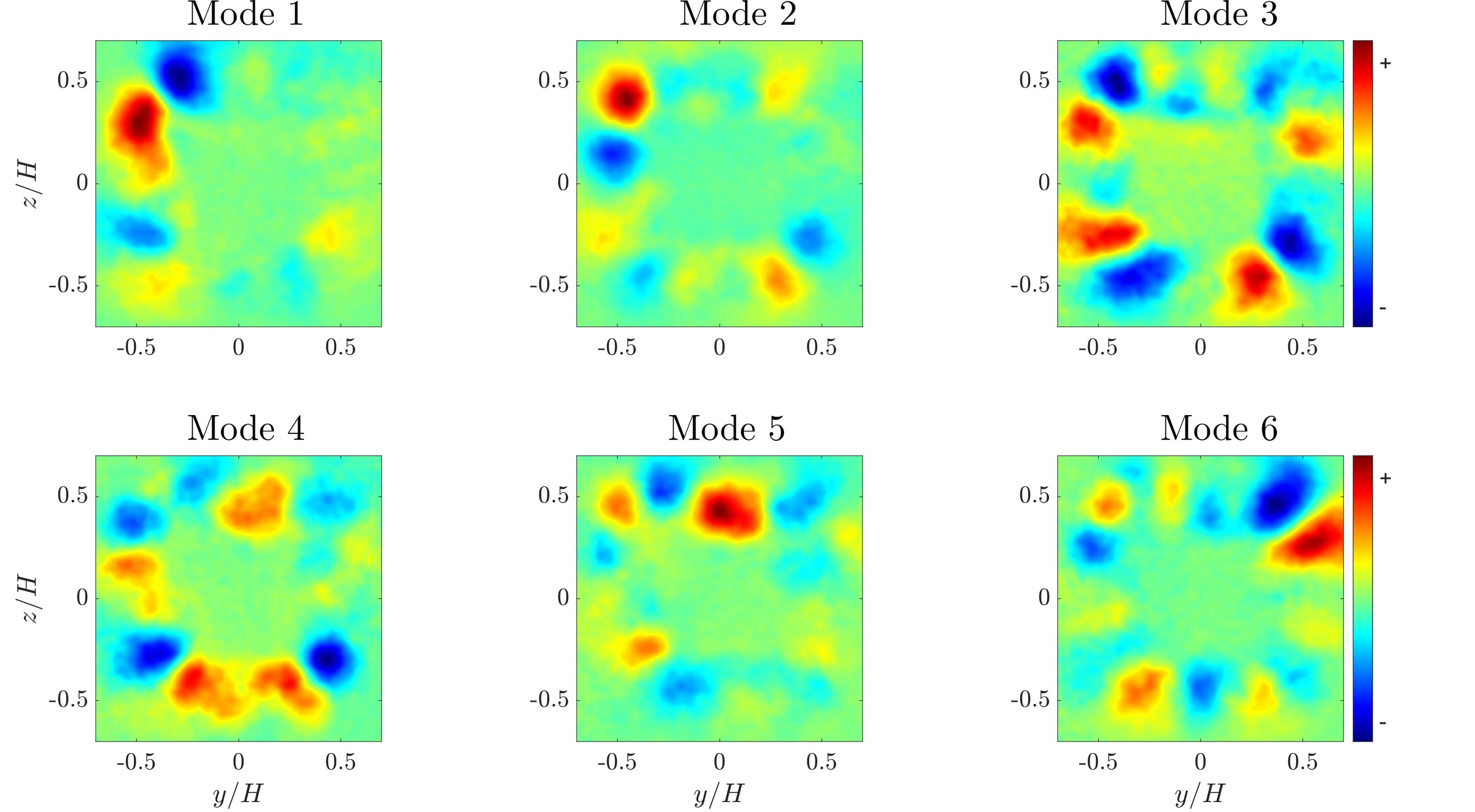}
		\caption{}
		\label{fig:POD_modes2H}
	\end{subfigure}

		\caption{Results of the snapshot POD from PIV measurements on the cross-stream plane $x=2H$ at 100\% PWM. (a) energy distribution of the leading POD modes; (b) streamwise components of the first 6 POD modes.}
	\label{fig:POD2H}
\end{figure}

In a similar manner, we apply the snapshot POD to investigate the dominant turbulent structures at two cross-stream measurement planes under 100\% PWM, and the results are presented in \Cref{fig:POD1H} and \Cref{fig:POD2H}. 
In both cases, the leading POD modes contain only fractional amounts of the overall TKE. 
Comparatively the energy convergence at $x=2H$ is slightly faster than at $x=1H$. 
In terms of the shapes of the dominant turbulent structures, the streamwise components of the leading POD modes at $x=1H$ are dominated by small-scaled turbulent structures that are non-uniformly distributed inside the shear layer region. 
These small-scaled structures progress downstream and interact with each other. 
At $x=2H$, the dominant turbulent structures become larger, and clear azimuthal patterns can be observed inside the leading POD modes, although the turbulent shear layer is not strictly axisymmetric.

\section{Conclusion and Outlook}\label{sec:conclusions}

In this study, 
we design and construct the Small Fan Array Wind Generator (SFAWG), 
featuring individual control of $10\times 10$ fan elements. 
We utilize uniform duty cycles for the fan array, and perform aerodynamic characterization of the facility using hotwire measurements in the near field ($\mathcal{O}(h)$), as well as Particle Image Velocimetry (PIV) measurements in one streamwise plane and 
two cross-stream planes in the far field ($\mathcal{O}(H)$).
The major observations are concluded as follows:
\begin{itemize}
 \item In the near field, the outer flow is dominated by the mixing process, 
 where high momentum ``swirling jets'' generated from the fans expand and influence the neighboring low speed region.  
 A nearly uniform mixing can be achieved at the end of the near field.
 \item In the far field, the flow characteristics are similar to those of a typical jet, 
 including the expansion of the jet shear layer and the decay of the core region. 
 Moreover, the annular shear layer is stretched and distorted as the flow progresses downstream, 
 which results in the multi-scaled, energy-dispersive turbulent structures.
\item The complex dynamics in both the near and far fields severely limit the flow region 
with the desired properties of uniformity and low turbulence level. 
At a downstream distance of $2H$, only about 36\% of the cross-sectional area has satisfying flow properties, 
and this ratio drops to about 25\% at $3H$ downstream.
	\item The flow profiles generated at different duty cycles exhibit strong similarity, 
	indicating that the challenges discussed above will exist under a wide range of duty cycles.
\end{itemize} 
Preliminary investigations yield similar observations 
for the world's largest FAWG for drone testing. 
This FAWG has a 3.25 m $\times$ 3.25m active blowing frame
accomodating 1600 individually controllable computer fans \cite{LiuYT2023FluCoMe}.
Hence, we hypothesize that our results as typical for all square FAWG.

The major conclusions of this study highlight the need to improve the flow generated from FAWGs. 
To improve the flow quality, mesh grids and honeycombs may be applied to enable a reduced turbulence level. 
However, the employment of such devices will unavoidably lead to energy loss, so a dedicated design of the flow straightening devices may help to achieve a balance between flow quality and energy loss.
To reduce the shear layer expansion, local flow control devices can be adopted,
like vortex generators for jets \cite{Zamam1994pf}.

The generation of more accurate spatial-temporal wind profiles
will benefit from general nonuniform operation of the fans with suitable control schemes\cite{Brunton2015amr}.
In the pioneering works of \citep{Ozono_2018,Wang_2019,Li_2021}, the control of turbulence level and wind profiles are achieved using various control algorithms.
With the implementation of upstream control and downstream sensing, 
the addition of state-of-the-art control strategies may further improve the flow profiles at desired downstream location.
The fan array wind generators possess undoubted potential to empower the design and test of future UAVs.


\section*{Funding Sources}

This work is supported 
by the National Science Foundation of China (NSFC) through grants 12172109, 12172111, and 12202121, 
by Guangdong province, China, 
by the Shenzhen Science and Technology Program under grant JCYJ20220531095605012, 
by the Natural Science and Engineering grant 2022A1515011492,  and 
by the Science and Technology Innovation Bureau, Pingshan District, Shenzhen City, China, through grant 29853M-KCJ-2023-002-05.

\section*{Acknowledgments}
We acknowledge valuable discussions 
with Jialong Chen,  Zhibin Chen, Guy Yoslan Cornejo Maceda, Nan Deng, Nan Gao, Fran\c{c}ois Lusseyran, Jincai Yang, Jun Yang, Yang Yang, Yannian Yang, Huang Yao and Yanlung Zhang.
We appreciate generous technical and scientific support from the HangHua company (Dalian, China).
We also extend our thanks to Prof. Xiangyuan Zheng and Prof. Sunwei Li from Tsinghua Shenzhen International Graduate School for their generous help during the experiments.
We are grateful to the anonymous referees for their insightful and constructive advice.
\bibliography{Main}

\end{document}